\documentclass[12pt]{article}
\usepackage{epsf,amsfonts,color,appendix}
\usepackage{graphicx}
\newcommand\encadremath[1]{\vbox{\hrule\hbox{\vrule\kern8pt
\vbox{\kern8pt \hbox{$\displaystyle #1$}\kern8pt}
\kern8pt\vrule}\hrule}}
\def\enca#1{\vbox{\hrule\hbox{
\vrule\kern8pt\vbox{\kern8pt \hbox{$\displaystyle #1$}
\kern8pt} \kern8pt\vrule}\hrule}}

\newcommand\figureframex[3]{
\begin{figure}[bth]
\hrule\hbox{\vrule\kern8pt
\vbox{\kern8pt \vbox{
\begin{center}
{\mbox{\epsfxsize=#1.truecm\epsfbox{#2}}}
\end{center}
\caption{#3}
}\kern8pt}
\kern8pt\vrule}\hrule
\end{figure}
}
\newcommand\figureframey[3]{
\begin{figure}[bth]
\hrule\hbox{\vrule\kern8pt
\vbox{\kern8pt \vbox{
\begin{center}
{\mbox{\epsfysize=#1.truecm\epsfbox{#2}}}
\end{center}
\caption{#3}
}\kern8pt}
\kern8pt\vrule}\hrule
\end{figure}
}

\makeatletter
\@addtoreset{equation}{section}
\makeatother

\newtheorem{theorem}{Theorem}[section]

\newtheorem{remark}{Remark}[section]
\newtheorem{proposition}{Proposition}[section]
\newtheorem{lemma}{Lemma}[section]
\newtheorem{corollary}{Corollary}[section]
\newtheorem{definition}{Definition}[section]
\def\br{\begin{remark}\rm\small}
\def\er{\end{remark}}
\def\bt{\begin{theorem}}
\def\et{\end{theorem}}
\def\bd{\begin{definition}}
\def\ed{\end{definition}}
\def\bp{\begin{proposition}}
\def\ep{\end{proposition}}
\def\bl{\begin{lemma}}
\def\el{\end{lemma}}
\def\bc{\begin{corollary}}
\def\ec{\end{corollary}}
\def\beaq{\begin{eqnarray}}
\def\eeaq{\end{eqnarray}}
\newcommand{\proof}[1]{{\noindent \bf proof:}\par
{#1} $\square$}

\newcommand{\beq}{\begin{equation}}
\newcommand{\eeq}{\end{equation}}
\newcommand{\bea}{\begin{eqnarray}}
\newcommand{\eea}{\end{eqnarray}}

\newcommand{\Res}{\mathop{\,\rm Res\,}}

\usepackage[pdftex]{hyperref}
\hypersetup{colorlinks,urlcolor=magenta,citecolor=red,linkcolor=blue,filecolor=black}

\begin{document}
\sloppy

\pagestyle{empty}
\begin{flushright}
IPhT-T10/166 \\ CERN-PH-TH/2010-265
\end{flushright}

\addtolength{\baselineskip}{0.20\baselineskip}
\begin{center}
\vspace{26pt}
{\large \bf {Tracy-Widom GUE law and symplectic invariants}}
\newline
\vspace{26pt}

{\sl G.\ Borot}\hspace*{0.05cm}\footnote{\href{mailto:gaetan.borot@cea.fr}{gaetan.borot@cea.fr}},
{\sl B.\ Eynard}\hspace*{0.05cm}\footnote{\href{mailto:bertrand.eynard@cea.fr}{bertrand.eynard@cea.fr}}

\vspace{6pt}
${}^{1}$ Institut de Physique Th\'{e}orique de Saclay,\\
F-91191 Gif-sur-Yvette Cedex, France \\

\vspace{0.1cm}

${}^{2}$ CERN, Theory Division \\
CH-1211 Geneva 23, Switzerland

\end{center}

\vspace{20pt}
\begin{center}
{\bf Abstract}
\end{center}

%

\vspace{0.5cm}
We establish the relation between two objects: an integrable system related to Painlev\'{e} II equation, and the symplectic invariants of a certain plane curve $\Sigma_{TW}$ describing the average eigenvalue density of a random hermitian matrix spectrum near a hard edge (a bound for its maximal eigenvalue). This explains directly how the Tracy-Widow law $\mathrm{F}_{\mathrm{GUE}}$, governing the distribution of the maximal eigenvalue in hermitian random matrices, can also be recovered from symplectic invariants.  \newpage


\vspace{26pt}
\pagestyle{plain}
\setcounter{page}{1}


\tableofcontents

\newpage

\section*{Motivation}

The famous Tracy-Widom law \cite{TW93}, describes the statistics of the largest eigenvalue of a random hermitian matrix of the GUE ensemble \cite{Mehtabook}.

Consider the large deviation function
\beq
\mathcal{F}_{N}(a) \equiv \ln \mathrm{Prob}[\lambda_{\mathrm{max}} \leq a]\qquad  \mathrm{for}\;a < 2
\eeq
Here $2 = \lim_{N \rightarrow \infty} \mathbb{E}(\lambda_{\max})$. It is proved, that $\mathcal{F}_N(a)$ has an expansion in $1/N^2$, either in \cite{Scher} by probabilistic methods, or in \cite{Erc} by Riemann-Hilbert asymptotic analysis for orthogonal polynomials \cite{BI1999,D1,D2,D3}:
\beq
\label{eq:Phia}\mathcal{F}_{N}(a) =  A_N + \sum_{g \geq 0} N^{2 - 2g}\,F^{g}(a)
\eeq
with $F^{g}(a)$ independent of $N$.
It is known that the $F^{g}(a)$ are the symplectic invariants introduced in \cite{EOFg} (see Section~\ref{sec:toporec} below). To be short, they can be computed recursively, purely by means of algebraic geometry on a certain plane curve $\Sigma(a)$.

In \cite{BEMN},
we have computed those coefficients $F^{g}(a)$, also for the genealization to
$\beta$ ensembles ($\beta > 0$) with polynomial potential. In this article, we shall consider only the hermitian case ($\beta = 2$). Because of universality, we may restrict ourselves to a quadratic potential, i.e. to the GUE ensemble.

Then, heuristically, one can study not so large deviations by pluging $a = 2 + N^{-2/3}s$ ($s < 0$) in Eqn.~\ref{eq:Phia}. The result is that (see \cite{BEMN}), with this scaling, each term of the expansion becomes of order $1$:
\beq
\label{eq:asym}\mathcal{F}_{N}(a = 2 + N^{-2/3}s) \sim A_{\mathrm{GUE}} + \sum_{g \geq 0} (-s/2)^{3(1 - g)}\,F^{g}(\Sigma_{\mathrm{TW}})
\eeq
$A_{\mathrm{GUE}}$ is a constant that one can compute, and $F^{g}(\Sigma_{\mathrm{TW}})$ are the symplectic invariants of a certain plane curve $\Sigma_{\mathrm{TW}}$, which is a limit of the curve $\Sigma(a)$ when $a \rightarrow 2$. This limit object is in fact universal (it is the same if we start from a generic potential), it is the curve of equation $y^2 = x + \frac{1}{x} - 2$, up to normalization of $x$ and $y$. Heuristically, it is expected that the RHS of Eqn.~\ref{eq:asym} is the exact $s \rightarrow -\infty$ asymptotic of $\lim_{N \rightarrow \infty} \mathcal{F}_{N}(a = 2 + N^{-2/3}s)$. Indeed, this method for $\beta > 0$ reproduces all previously known results \cite{DM06,RRV,DIK,BBDF07} on the left tail asymptotics of $\beta$ Tracy-Widom laws. It also predicts the constant for any $\beta > 0$, and provides an algorithm to compute the expansion at all orders. In our case, for GUE:
\beq
\label{eq:PGUE}\mathcal{P}^*_{\mathrm{GUE}}(s) = 2^{1/24}e^{\zeta'(-1)}\,\exp\left(-\frac{|s|^3}{12} - \frac{\ln|s|}{8} + \sum_{g \geq 2} (-s/2)^{3(1 - g)}\,F^{g}(\Sigma_{\mathrm{TW}})\right)
\eeq
From Tracy and Widom \cite{TW93}, we know that $\mathcal{P}_{\mathrm{GUE}}^*(s) = \mathrm{F}_{\mathrm{GUE}}(s)$, where $\mathrm{F}_{\mathrm{GUE}}$ is related to Painlev\'{e} II equation, which itself appears in relation to an integrable system. However, by performing a $1/N$ expansion and computing $F^{g}$, the integrability structure is not manifest. At first, it was not clear why the asymptotic series in the RHS of Eqn.~\ref{eq:PGUE} should be related to Painlev\'{e} II. In this article, we fill this gap and give a proof of the following:
\begin{proposition}
\label{prop1}
\beq
\exp\left(-\frac{|s|^3}{12} - \frac{\ln|s|}{8} + \sum_{g \geq 2} (-s/2)^{3(1- g)}\,F^g(\Sigma_{\mathrm{TW}})\right)
\eeq
is the asymptotic when $s \rightarrow -\infty$ of a Tau function for two $2\times 2$ compatible systems of ODE:
\beq
\partial_x \Psi(x,s) = \mathbf{L}(x,s)\Psi(x,s),\qquad \partial_s \Psi(x,s) = \mathbf{M}(x,s)\Psi(x,s)
\eeq
whose compatibility condition is equivalent to a Painlev\'{e} II equation:
\beq
q''(s) = 2q^3(s) + sq(s)
\eeq
\end{proposition}
The identity of Eqn.~\ref{eq:asym} was based on a heuristic argument, the double scaling limit of a matrix model, which would require a mathematical proof. In this article, we start from the RHS given by heuristics (i.e. from the world of the topological recursion), and we prove that it is indeed the expansion of the LHS. We do not use a Fredholm determinant representation of $\mathrm{Prob}[\lambda_{\mathrm{max}} \leq a]$.

\newpage

\section*{Outline}

In the first part of the article, we gather older facts relevant for our problem:
\begin{itemize}
\item[$\bullet$] the framework of the topological recursion and its properties (Section~\ref{sec:toporec}-\ref{sec:M}).
\item[$\bullet$] the general relation between $2\times 2$ integrable systems and loop equations (Section~\ref{sec:loopeqint}-\ref{sec:N}).
\end{itemize}

In a second part, we refine the technical hypothesis underlying this very relation (the outcome is Theorem~\ref{th3}, our main theoretical result). We explain what really has to be checked (Section~\ref{sec:hypo3}), and we sketch how this be done in practice (Section~\ref{sec:ana}).

The third part is devoted to the application of these ideas to an integrable system related to Painlev\'{e} II (Sections~\ref{sec:intp2}-\ref{sec:AAA}). Subsequently, we prove Prop.~\ref{prop1} (Sections~\ref{sec:1N}-\ref{sec:Con}).

Let us mention that many ideas used in the proof also appeared in the work of M.~Cafasso and O.~Marchal \cite{CM10} with different purposes (they related the $(2m,1)$ minimal models appearing in the merging of two cuts in the 1 hermitian matrix model, to a hierarchy of equations containing Painlev\'{e} II), and in fact have their origin in the works of the second author with M.~Berg\`{e}re \cite{BE09,BE09l}. In this article, our goal is to close (in the hermitian case only) the alternative approach to Tracy-Widom laws presented in \cite{BEMN}.

\vspace{1cm}

\section*{Acknowledgments}

We would like to thank M.~Berg\`{e}re for sharing his computations on Schr\"{o}dinger equations and frequent valuable discussions, T.~Grava for discussions about PII, O.~Marchal and A.~Its for comments, S.N.~Majumdar, C.~Nadal with whom we collaborated on a previous related work, and G.~Schehr for motivating discussions. G.B. thanks the hospitality of the SISSA and of the Department of Maths and Statistics of Melbourne University, where part of this work was completed, and the organizers of the MSRI semester on Random Matrix Theory. The work of B.E. is partly supported by the ANR project GranMa "Grandes Matrices Al\'{e}atoires" ANR-08-BLAN-0311-01, by the European Science Foundation through the Misgam program, by the Qu\'{e}bec government with the FQRNT. He also would like to thank the CRM (Centre de recherches math\'ematiques de Montr\'eal, QC, Canada) and the CERN for its hospitality.

\newpage

\section{Basics around the topological recursion}
\label{sec:bas}
\subsection{Definitions}
\label{sec:toporec}

The topological recursion and symplectic invariants were axiomatically defined in \cite{EOFg}, and we refer to \cite{EORev} for a review, and to Appendix~\ref{appB} for a summarized definition. It consists in an algorithm associating some numbers $F^{g}(\Sigma)$ and differential forms $\omega_n^{g}(\Sigma)(z_1,\ldots,z_n)$ to a regular spectral curve $\Sigma$.

For our purposes, we call \emph{spectral curve} the data of:
\begin{itemize}
\item[$\bullet$] a plane curve $(\mathcal{C},x,y)$, in other words a Riemann surface $\mathcal{C}$, with two meromorphic functions $x$ and $y$, $\mathcal{C}\mapsto \mathbb C$.
\item[$\bullet$] a maximal open domain $U \subseteq \mathcal{C}$ on which $x$ is a coordinate patch, called \emph{physical sheet}.
\item[$\bullet$] a Bergman kernel $B(z_1,z_2)$, i.e. a differential form in $z_1 \in \mathcal{C}$ and in $z_2 \in \mathcal{C}$, such that, in any local coordinate $\xi$:
\beq
B(z_1,z_2)  \mathop{=}_{z_1 \rightarrow z_2} \frac{\mathrm{d}\xi(z_1)\mathrm{d}\xi(z_2)}{(\xi(z_1) - \xi(z_2))^2} + O(1)
\eeq
\end{itemize}
The zeroes of $\mathrm{d}x$ 
are called \emph{branchpoints} (name them $a_i$), and a spectral curve is said to be \emph{regular} when these zeroes are simple and are not zeroes of $\mathrm{d}y$. In other words, when $\sqrt{x - x(a_i)}$ is a good coordinate around $a_i \in \mathcal{C}$, and when $y$ behaves like $y(z)\sim y(a_i)+y'(a_i)\,\sqrt{x(z)-x(a_i)}$ with $y'(a_i)\neq 0$.

We shall not give the full definitions of $F^{g}(\Sigma) = \omega_0^{g}(\Sigma)$ and $\omega_n^{g}(\Sigma)$ here, and rather refer to Appendix~\ref{appB} or \cite{EOFg}.  Their construction is axiomatic and relies only on algebraic geometry on $\mathcal{C}$. The construction is made by recursion on $2g-2+n$.

We rather mention the essential properties that we use here:
\begin{itemize}
\item[$\bullet$] $F^{g}(\Sigma)$ is invariant under any transformation $(x,y) \rightarrow (x_o,y_o)$ such that $|\mathrm{d}x\wedge\mathrm{d}y| = |\mathrm{d}x_o\wedge\mathrm{d}y_o|$. For this reason, the $F^{g}(\Sigma)$ are called \emph{symplectic invariants}.
\item[$\bullet$] For $2 - 2g - n < 0$, $\omega_n^{g}(\Sigma)\in T^*({\cal C})\otimes \dots \otimes T^*({\cal C})$, i.e. $\omega_n^{g}(\Sigma)(z_1,\ldots,z_n)$ is a meromorphic differential form in each $z_i \in \mathcal{C}$, symmetric in all $z_i$'s, and it has poles only at the 
branchpoints
, of maximal degree $2(3g + n - 2)$, with vanishing residue.
\item[$\bullet$] for $2-2g-n<0$, $\omega_n^{g}(\Sigma)$ are homogeneous of degree $2-2g-n$, i.e. if we change $y\to \lambda y$, we have $\omega_n^{g}(\Sigma) \to \lambda^{2-2g-n}\,\omega_n^{g}(\Sigma)$, and $F^{g}(\Sigma) \to \lambda^{2-2g}\,F^{g}(\Sigma)$.
\item[$\bullet$] $\omega_n^{g}(\Sigma)$ have nice scaling properties when the spectral curve approaches a singular spectral curve. We will be more precise when needed.
\item[$\bullet$] $\omega_n^{g}(\Sigma)$ have nice properties under variation of the spectral curve. If $\Sigma_t$ is a 1-parameter family of spectral curves, let us write $(\partial_{t} y\, \mathrm{d}x - \partial_{t} x\, \mathrm{d}y) = \Omega$, and represent the differential form  $\Omega$ with the Bergman kernel:
    \beq
    \Omega(z) = \int_{\Omega^*} \Lambda_{\Omega}(\xi)B(\xi,z)
    \eeq
(this is form-cycle duality: $\Omega^*$ is the cycle dual to the differential form $\Omega$, and in case the cycle is not regular, it is accompanied by a Jacobian $\Lambda_{\Omega}$).
    Then:
    \beq
    \label{eq:varF}\partial_t \omega_n^{g}(\Sigma_t)(z_1,\ldots,z_n) = \int_{\Omega^*} \Lambda_{\Omega}(\xi)\,\omega_{n + 1}^{g}(\Sigma_t)(\xi,z_1,\ldots,z_n).
    \eeq
\end{itemize}

\subsection{Matrix models and topological recursion}
\label{sec:M}
The topological recursion allows to find all order asymptotics of the large $N$ expansion of matrix models. Let us recall how it works for the 1 hermitian matrix model with one hard edge, relevant for the application to Tracy-Widom law. For any potential $V$ such that $|\int_{\mathbf{R}} \mathrm{d}\xi e^{-\epsilon V(\xi)}| < \infty$ for some positive $\epsilon$, let us consider $N$ real random variables $\lambda_1,\ldots,\lambda_N$ with joint probability law of density:
\beq
\mathrm{d}\mu_{N}(\lambda) = \frac{1}{Z_{N}(\infty)}\prod_{1 \leq i < j \leq N} |\lambda_i - \lambda_j|^{2} \cdot \prod_{i = 1}^N \mathrm{d}\lambda_i\,\exp\left(-N V(\lambda_i)\right)
\eeq
The probability of having $\textrm{max}\,\lambda_i \leq a$ is given by $\mathcal{P}(a) = Z_{N}(\left.]-\infty,a\right.])/Z_{N}(\infty)$, where:
\beq
Z_{N}(J) = \int_{J^N} \mathrm{d}\mu_{N}(\lambda)
\eeq
$\mathrm{d}\mu_{N}$ is the probability distribution induced on the eigenvalues of a random hermitian matrix of size $N \times N$, with probability law $\mathrm{d}M\,\exp\left(-\frac{N}{t} \mathrm{Tr}\,V(M)\right)$ up to normalization. $\mathrm{d}M$ here is the canonical Lebesgue measure on the real vector spaces of hermitian matrices. The observables invariant under conjugation are spanned by the cumulants:
\beq
W_n(x_1,\ldots,x_n) = \Big\langle \prod_{i = n}^N \left(\sum_{j_i = 1}^N \,\frac{1}{x_i - \lambda_{j_i}}\right)\Big\rangle_c
\eeq
When $\ln Z_{N}(a)$ admits an asymptotic expansion in powers of $1/N^2$,
we write:
\bea
Z_{N}(a) & = & C_N\,\exp\left(\sum_{g \geq 0} N^{2 - 2g}\,F^{g}\right) \\
W_n(x_1,\ldots,x_n) & = & \sum_{g \geq 0} N^{2 - 2g - n}\,W_n^{g}(x_1,\ldots,x_n)
\eea
where $C_N$ is a normalization constant depending only on $N$, and $F^{g}$ and $W_n^{g}$ are independent of $N$ but depend on $a$ and $V$. Before coming to the coefficients themselves, let us say that when the density of eigenvalues has a large $N$ limit, $\rho(\xi)\mathrm{d}\xi$, which admits a single interval $J_0 \subseteq ]-\infty,a[$ as support, it is proved that such an expansion exists \cite{Scher,Erc}. If the support consists of several intervals, no such expansion exists (there are fast oscillatory terms depending on $N$), but there is still a way to extract the partition function from the topological recursion \cite{Eyn2008}. Yet, this discussion is out of the scope of this article.
The theorem proved in \cite{Eyn2004,CE06} is that when the large $N$ expansion exists, then:
\bea
F^{g} & = &  F^{g}(\Sigma(a)) \\
W_n^{g,k}(x(z_1),\ldots,x(z_n))\mathrm{d}x(z_1)\cdots\mathrm{d}x(z_n) & = & \omega_n^{g,k}(\Sigma(a))(z_1,\ldots,z_n)
\eea
for the spectral curve $\Sigma(a)$ determined by the equation $y = V'(x)/2 - W_1^{0}(x)$ and the Bergman kernel determined by:
\beq
B(z_1,z_2) = \left(W_2^{0}(x(z_1),x(z_2)) + \frac{1}{(x(z_1) - x(z_2))^2}\right)\mathrm{d}x(z_1)\mathrm{d}x(z_2)
\eeq
$y(x)$ is analytic on the $x$ complex plane, except on the eigenvalue support $J_0$, where it admits a discontinuity $y(x + i0^+) - y(x - i0^-) = 2i\pi\rho(x)$. $\mathcal{C}$ is a Riemann surface on which $y$ can be analytically continued.

\subsection{$2\times 2$ differential system and topological recursion}
\label{sec:loopeqint}

To any $2\times 2$ first order differential system:
\beq
\label{eq:psipsi}\partial_x \Psi = \mathbf{L}\Psi,\qquad  \Psi = \left(\begin{array}{cc} \psi & \phi \\ \overline{\psi} & \overline{\phi}\end{array}\right)
\eeq
is associated the integrable kernel:
\beq
\mathcal{K}(x_1,x_2) = \frac{\psi(x_1)\overline{\phi}(x_2) - \overline{\psi}(x_1)\phi(x_2)}{x_1 - x_2}
\eeq
We shall restrict ourselves to $\mathrm{Tr}\,\mathbf{L} = 0$. So, $\partial_x\big(\mathrm{det}\,\Psi\big) = 0$, and we can choose the normalization:
\beq
\label{eq:norm}\mathrm{det}\,\Psi = 1
\eeq
In \cite{BE09} were also introduced correlators $\overline{\mathcal{W}}_n(x_1,\ldots,x_n)$ and connected correlators $\mathcal{W}_n(x_1,\ldots,x_n)$. The connected correlators are defined by:
\bea
\mathcal{W}_1(x) & = & \lim_{x' \rightarrow x}\left(\mathcal{K}(x,x') - \frac{1}{x - x'}\right)   \\
\mathcal{W}_2(x_1,x_2) & = & - \mathcal{K}(x_1,x_2)\mathcal{K}(x_2,x_1) - \frac{1}{(x_1 - x_2)^2}  \\
\label{eq:dee}\mathcal{W}_n(x_1,\ldots,x_n) & = & (-1)^{n + 1} \sum_{\sigma\,\mathrm{cycles}\,\mathrm{of}\,\mathfrak{S}_n} \prod_{i = 1}^n \mathcal{K}(x_i,x_{\sigma(i)})
\eea
and the correlators by:
\beq
\overline{\mathcal{W}}_n(x_1,\ldots,x_n) = "\mathrm{det}"\,\mathcal{K}(x_i,x_j)
\eeq
where "det" means that each occurence of $\mathcal{K}(x_i,x_i)$ in the determinant should be replaced by $\mathcal{W}_1(x_i)$, and each occurence of $\mathcal{K}(x_i,x_j)\mathcal{K}(x_j,x_i)$ by $-\mathcal{W}_2(x_i,x_j)$.
In other words the ${\mathcal{W}}_n$ are the cumulants of the $\overline{\mathcal{W}}_n$.

For example:
\beq
\mathcal{W}_1  = \psi\,\partial_x\overline{\phi} - \overline{\psi}\,\partial_x \phi = -\big(\partial_x\psi\,\,\overline{\phi} - \partial_x\overline{\psi}\,\,\phi\big)
\eeq
Eqn.~\ref{eq:norm} implies that all correlators are symmetric in the $x_i$'s. It can be checked that they do not have poles at coinciding points $x_i = x_j$, $i \neq j$.
The spectral curve of a first order differential system is defined by the plane curve $\mathcal{C}$ of equation:
\beq\label{eqdefspcurveL}
\label{eq:sp}\mathrm{det}(y - \mathbf{L}(x)) =  0 .
\eeq
\begin{theorem} (proved in \cite{BE09})
\label{th2}
Assume that:
\begin{itemize}
\item[$(i)$] $\mathbf{L}$ depends on some parameter $N$, and has a limit when $N \rightarrow \infty$.
\item[$(ii)$] The spectral curve $\Sigma_N$ of the system Eqn.~\ref{eqdefspcurveL} has a large $N$ limit $\Sigma_{\infty}$ which is regular, and has genus $0$.
\item[$(iii)$] $\mathcal{W}_n(x_1,\ldots,x_n)$ admits an asymptotic expansion when $N \rightarrow \infty$ of the form $\mathcal{W}_n = \sum_{g \geq 0} N^{1 - 2g}\,\mathcal{W}_n^{g}$, and for $2g-2+n>0$, $\mathcal{W}_n^{g}(x_1,\ldots,x_n)$ may have singularities only at branchpoints of $\Sigma_{\infty}$.
\end{itemize}
Then, the expansion coefficients of the correlators are computed by the topological recursion applied to the spectral curve defined by $\Sigma_{\infty}$ with Bergman kernel $B(z_1,z_2) = (\mathcal{W}_2^{0}(x(z_1),x(z_2))+1/(x(z_1)-x(z_2))^2)\mathrm{d}x(z_1)\mathrm{d}x(z_2)$:
\beq
\mathcal{W}_n^{g}(x(z_1),\ldots,x(z_n))\mathrm{d}x(z_1)\cdots\mathrm{d}x(z_n) = \omega_n^{g}(\Sigma_{\infty})(z_1,\ldots,z_n)
\eeq
\end{theorem}
We stress that it is the large $N$ limit of the spectral curve (also called "classical", or "semiclassical" spectral curve) which is relevant in this theorem. A similar result holds when $\Sigma_{\infty}$ is not of genus $0$ under an extra hypothesis, but this will not be needed here.
So far, we did not speak of an integrable system, the theorem itself tells something about one differential system alone. However, in many examples and in this article, proving that (iii) holds is made possible by the existence of other compatible systems $\partial_{t_j} \Psi = \mathbf{M}_j \Psi$, where $t_j$ is a parameter of $\mathbf{L}$, and $\mathbf{M}_j$ is rational in $x$. So, we shall see in Sections~\ref{sec:hypo3}-\ref{sec:BJ} that hypothesis $(iii)$ can be considerably weakened.

\subsection{Generalities on the Schr\"{o}dinger equation}
\label{sec:Berg}
Let us write:
\beq
\mathbf{L} = \left(\begin{array}{cc} a & b \\ c & -a \end{array}\right)
\eeq
Eqn.~\ref{eq:psipsi} imply that $\psi$ and $\phi$ are two independent solutions of a second order equation:
\beq
\label{efs} \partial_x^2 \psi + B\,\partial_x \psi + V\,\psi = 0
\eeq
where:
\bea
B & = & -\frac{\partial_x b}{b}  \\
V & = & \mathrm{det}\,\mathbf{L} - \partial_x a + a\frac{\partial_x b}{b}
\eea
All the same, $\overline{\psi}$ and $\overline{\phi}$ are independent solutions of the second order equation:
\beq
\label{efs2} \partial_x^2 \overline{\psi} + \overline{B}\,\partial_x \psi + \overline{U}\,\psi = 0
\eeq
where:
\bea
\overline{B} & = & - \frac{\partial_x c}{c} \\
\overline{V} & = & \mathrm{det}\,\mathbf{L} + \partial_x a - a \frac{\partial_x c}{c}
\eea
We shall use a classical change of function in the theory of Schr\"{o}dinger equations to study the asymptotics of $\Psi$ at singularities of $\mathbf{L}$. We are grateful to M.~Berg\`{e}re who shared his experience on that subject with us. We can always write:
\bea
\label{eq:says1}\psi & = & \mathrm{cte}\,\sqrt{\frac{bf}{2}}\exp\left(-\int^x \frac{\mathrm{d}\xi}{f(\xi)}\right) \\
\label{eq:says2}\overline{\psi} & = & \overline{\mathrm{cte}}\,\sqrt{\frac{c\overline{f}}{2}}\exp\left(-\int^x \frac{\mathrm{d}\xi}{\overline f(\xi)}\right)
\eea
Then, Eqns.~\ref{efs}-\ref{efs2} are equivalent to:
\bea
\frac{1}{2}\partial_x^3\,f & = & 2U(\partial_x f) + (\partial_x U)f  \\
\frac{1}{2}\partial_x^3\,\overline{f} & = & 2\overline{U}(\partial_x \overline{f}) + (\partial_x \overline{U})\overline{f}
\eea
with $U,\overline{U}$ given by $-V,-\overline{V}$ up to a schwartzian derivative:
\bea
U & = & -V + \frac{3}{4}\frac{(\partial_xb)^2}{b^2} - \frac{1}{2}\frac{\partial_x^2 b}{b}  \\
\overline{U} & = & - \overline{V} + \frac{3}{4}\frac{(\partial_xc)^2}{c^2} - \frac{1}{2}\frac{\partial_x^2 c}{c}
\eea

\subsection{Tau function and topological recursion}
\label{sec:N}
Here, we modify slightly the systems by introducing a parameter $1/N$ in front of each derivative (very often, $N$ turns out to be redundant with other parameters). For a large class of $2\times 2$ first order compatible systems (for example this includes the $(p,2)$ minimal models which are finite reductions of KdV) of the type:
\beq
\label{eq:int}\frac{1}{N}\partial_x \Psi = \mathbf{L}\Psi\,\qquad \frac{1}{N}\partial_{t_j} \Psi = \mathbf{M}_j\Psi
\eeq
and under the hypothesis of Thm.~\ref{th2}, it was stated in \cite{BE09l,CM10} without details that $\exp\left(\sum_{g \geq 0} N^{2 - 2g}\,F^{g}(\Sigma_{\infty})\right)$ is a formal Tau function. We shall make the argument explicit here. We are considering \emph{any} system like Eqn.~\ref{eq:int}, such that $\mathbf{L}$ and $\mathbf{M}_j$ are rational fractions in $x$. We assume that $\mathbf{L}$ has a large $N$ limit $\mathbf{L}^{(0)}$, and an expansion in powers of $1/N$.


\subsubsection{Definition of the $\tau$-function}

Here, we consider that $N$ is a fixed parameter. To any solution $\Psi(x,\vec{t})$ of Eqn.~\ref{eq:int}, Jimbo-Miwa-Ueno \cite{JM81} associate a Tau function $\tau(\vec{t})$ as follows. Let us consider the behavior of $\Psi(x,\vec{t})$ near each pole $x_{\alpha}$ of $\mathbf{L}(x,\vec{t})$, and write the $x\to x_\alpha$ asymptotics (valid in some angular sector near $x_\alpha$):
\beq
\Psi(x,\vec{t}) = \widetilde{\Psi}_{\alpha}(x,\vec{t})\,\exp\big(\mathbf{T}_{\alpha}(x,\vec{t})\big)
\eeq
where $e^{\mathbf{T}_{\alpha}}$ contains the essential singularity and the pole of $\Psi(x,\vec{t})$, and  $\widetilde{\Psi}_{\alpha}(x,\vec{t})$ is analytical when $x \rightarrow x_{\alpha}$.
Then, the compatibility of the system Eqn.~\ref{eq:int} implies that
there exists a function $\tau(\vec{t})$ such that:
\bea
\partial_{t_j}\ln\tau & = & - \sum_{\alpha} \Res_{x \rightarrow x_{\alpha}} \mathrm{d}x\,\mathrm{Tr}\,\widetilde{\Psi}_{\alpha}^{-1}\big(\partial_x \widetilde{\Psi}_{\alpha}\big)\,\big(\partial_{t_j} \mathbf{T}_{\alpha}\big)  \\
\label{eq:SJ} & = & - \sum_{\alpha} \Res_{x \rightarrow x_{\alpha}} \mathrm{d}x\,\mathrm{Tr}\,\big(\Psi^{-1}\partial_x \Psi\big)\,e^{-\mathbf{T}_{\alpha}}\big(\partial_{t_j} \mathbf{T}_{\alpha}\big)e^{\mathbf{T}_{\alpha}}
\eea

\begin{proposition}
Assume that $\mathbf{T}_{\alpha}(x,\vec{t})$ is family of diagonal matrices, and write $\mathbf{T}_{\alpha} = T_{\alpha}\,\mathbf{\sigma_3} + c\,\mathbf{1}$ (where $\sigma_3={\rm diag}(1,-1)$). We have:
\beq
\label{eq:tau0}\partial_{t_j} \ln\tau = 2\sum_{\alpha} \Res_{x \rightarrow x_{\alpha}} \mathrm{d}x\, \big(\partial_{t_j}T_{\alpha}\big)\,\mathcal{W}_1
\eeq
with the correlator $\mathcal{W}_1(x) = (\partial_x\psi)\overline{\phi} - (\partial_x\overline{\psi})\phi$ introduced in Section~\ref{sec:loopeqint}.
\end{proposition}

\proof{$\Psi^{-1}\partial_x \Psi = \Psi^{-1} \mathbf{L} \Psi$ is traceless like $\mathbf{L}$, and $\mathbf{T}_{\alpha}(x,\vec{t})$ is a family of commuting matrices. Thus, only the component of $\mathbf{T}$ of the Pauli matrix $\mathbf{\sigma}_3$ is relevant. Eqn.~\ref{eq:SJ} yields:
\beq
\partial_{t_j} \ln \tau = -2 \sum_{\alpha} \Res_{x \rightarrow x_{\alpha}} \mathrm{d}x\,\big(\partial_{t_j} T_{\alpha}\big) \, (\Psi^{-1}\partial_x\Psi)_{11}
\eeq
And we compute using $\mathrm{det}\,\Psi = 1$:
\bea
\Psi^{-1}\,\partial_x\Psi  & = & \left(\begin{array}{cc}  \overline{\phi} & - \phi  \\ - \overline{\psi} & \psi \end{array}\right)\cdot\left(\begin{array}{cc} \partial_x\psi & \partial_x\phi \\ \partial_x\overline{\psi} & \partial_x\overline{\phi}\end{array}\right)  \\
& = & \left(\begin{array}{cc} \partial_x\psi\,\overline{\phi} - \partial_x\overline{\psi}\,\phi & \overline{\phi}\,\partial_x\phi - \phi\,\partial_x\overline{\phi} \\ \psi(\partial_x\overline{\psi}) - \overline{\psi}\,\partial_x\psi & -\overline{\psi}\,\partial_x\phi + \psi\,\partial_x\overline{\phi}\end{array}\right)  \\
& = & \left(\begin{array}{cc} -\mathcal{W}_1 & \overline{\phi}\,\partial_x\phi - \phi\,\partial_x\overline{\phi} \\ \psi\,\partial_x\overline{\psi} - \overline{\psi}\,\partial_x\psi & \mathcal{W}_1\end{array} \right)
\eea}

\subsubsection{Large $N$ limit}
\label{sec:lN}
We use Section~\ref{sec:Berg} and keep the same notations, except that now, each derivative comes with a power of $1/N$. This modification can be easily traced back:
\bea
\psi(x) & = & \mathrm{cte}\,\sqrt{\frac{b(x)f(x)}{2}}\exp\left(-N \int^x \frac{\mathrm{d}\xi}{f(\xi)}\right)  \\
\overline{\psi}(x) & = & \overline{\mathrm{cte}}\,\sqrt{\frac{c(x)\overline{f}(x)}{2}}\exp\left(-N\int^x \frac{\mathrm{d}\xi}{\overline{f}(\xi)}\right)
\eea
where $f$ and $\overline{f}$ satisfy:
\bea
\label{eq:diff1}\frac{1}{2N^2}\,\partial_x^3 f & = & 2U(\partial_x f) + (\partial_x U)f  \\
\label{eq:diff2} \frac{1}{2N^2}\,\partial_x^3\,\overline{f} & = & 2\overline{U}(\partial_x \overline{f}) + (\partial_x \overline{U})\overline{f}
\eea
with the potential:
\bea
\label{eq:pot1} U & = & -\mathrm{det}\,\mathbf{L} + \frac{1}{N}\left(\partial_x a - a \frac{\partial_x b}{b}\right) + \frac{1}{N^2}\left(\frac{3}{4}\frac{(\partial_x b)^2}{b^2} - \frac{1}{2}\frac{\partial_x^2 b}{b}\right) \\
\overline{U} & = & - \mathrm{det}\,\mathbf{L} + \frac{1}{N}\left(-\partial_x a + a\frac{\partial_x c}{c}\right) + \frac{1}{N^2}\left(\frac{3}{4}\frac{(\partial_x c)^2}{c^2} - \frac{1}{2}\frac{\partial_x^2 c}{c}\right)
\eea
Let us write with superscript $^0$ the large $N$ limit of these quantities. In particular:
\beq
\mathbf{L}^{(0)} = \left(\begin{array}{cc} a^{(0)} & b^{(0)} \\ c^{(0)} & -a^{(0)}\end{array}\right)
\eeq
Recall that we defined the spectral curve through the equation $\mathrm{det}(y - \mathbf{L}(x)) = 0$. Since $\mathbf{L}$ is traceless, it is equivalent to $y^2 = -\mathrm{det}\,\mathbf{L}(x)$. Hence, the large $N$ limit of the spectral curve has equation:
\beq
\label{eq:spA}\Sigma_{\infty}(\vec{t})\,:\qquad  y^2 = -\mathrm{det}\,\mathbf{L}^{(0)}(x)
\eeq
Let us define $y(x)$ by this equation, with a choice of the branch of the square root depending on the asymptotic $x \rightarrow x_{\alpha}$ we require to define $\psi$ (see Eqn.~\ref{eq:psi1} below). According to Eqns.~\ref{eq:pot1}, the potential in the large $N$ limit is given by:
\beq
U^{(0)}(x) = \overline{U}^{(0)}(x) =  y^2(x)
\eeq
And, from Eqn.~\ref{eq:diff1}, we get:
\beq
f^{(0)}(x) \propto 1/y(x),\qquad \overline{f}^{(0)}(x) \propto 1/y(x)
\eeq
Matching the asymptotics gives the proportionality constants and we get:
\beq
\label{eq:psi1}\psi^{(0)}(x) = \mathrm{cte}\,\sqrt{\frac{b^{(0)}(x)}{2y(x)}}\exp\left(-N \int^x y(\xi)\mathrm{d}\xi\right) ,
\eeq
\beq
\overline{\phi}^{(0)}(x) = \overline{\mathrm{cte}}\,\sqrt{\frac{c^{(0)}(x)}{2y(x)}}\exp\left(N \int^x y(\xi)\mathrm{d}\xi\right) .
\eeq

\subsubsection{$1/N$ expansion of the $\tau$-function}

We now want to study asymptotics when $N \rightarrow \infty$. We assume that
\beq
\mathbf{T}_{\alpha}(x) = N\,\mathbf{T}_{\alpha}^{(0)}(x) = -N\,\mathbf{\sigma}_3\,\,\left(\int^x_o y(\xi)\mathrm{d}\xi\,\right)_{-,\alpha}
\eeq
where $(\cdots)_{-,\alpha}$ is the divergent part when $x \rightarrow x_{\alpha}$, and $o$ is an arbitrary origin of integration. In other words, we assume that $\mathbf{L}(x)$ and $\mathbf{L}^{(0)}(x)$ have the same poles $\{x_{\alpha}\}$, that the behavior of $\Psi$ at these poles is entirely given by the large $N$ limit of the system. This happens quite often in practice.
Then, we rewrite Eqn.~\ref{eq:tau0}:
\beq
\label{eq:tau0A}\partial_{t_j} \tau =  - 2N \sum_{\alpha} \Res_{x \rightarrow x_{\alpha}} \left(\int^x_o \partial_{t_j} y(\xi)\mathrm{d}\xi\right)\mathcal{W}_1(x)
\eeq
Provided the hypothesis of Thm.~\ref{th2} hold, the topological recursion can be applied, with the spectral curve $(\Sigma_{\infty}(\vec{t}))$ defined in Eqn.~\ref{eq:spA}. Let us compare to $\mathcal{F}(\vec{t})$ defined by the large $N$ asymptotic series:
\beq
\mathcal{F}(\vec{t}) \equiv \sum_{g \geq 0} N^{2 - 2g}\,F^{g}(\Sigma_{\infty}(\vec{t}))
\eeq
We also define:
\beq
W_1(x) \equiv \sum_{g \geq 0} N^{1 - 2g}\,W_1^{g}(\Sigma_{\infty}(\vec{t}))(x)
\eeq
Let us call $\mathcal{C}$, the underlying curve of $\Sigma_{\infty}(\vec{t})$, and $B$ its Bergman kernel. We shall use the properties of the topological recursion (Section~\ref{sec:toporec}) to compute the variations of $\mathcal{F}(\vec{t})$ wrt $t_j$. We have to represent
\beq
(\partial_{t_j}y\,\mathrm{d}x)(\xi) = \Res_{z \rightarrow \xi} B(z,\xi)\int^{z}_o y\,\mathrm{d}x
\eeq
Notice that the RHS do not depend on $o$. For $2\times 2$ systems, $\Sigma_{\infty}(\vec{t})$ is hyperelliptic. So, each pole $x_{\alpha}$ of $\int^{z}_o y\mathrm{d}x$ has two preimages $z_{\alpha}$ and $\overline{z}_{\alpha}$ in $\mathcal{C}$. One can move the contour to surround these points:
\beq
(\partial_{t_j}y\,\mathrm{d}x)(\xi) = - \sum_{\alpha} \Res_{z \rightarrow z_{\alpha},\overline{z}_{\alpha}} B(z,\xi)\int^{z}_o \partial_{t_j} y\,\mathrm{d}x
\eeq
Subsequently:
\beq
\partial_{t_j} F^{g}(\Sigma_{\infty}(\vec{t})) = - \sum_{\alpha}\Res_{z \rightarrow z_{\alpha},\overline{z}_{\alpha}} \mathrm{d}x(z)\,W_1^{g}(z)\int^z_o \partial_{t_j} y\,\mathrm{d}x
\eeq
In the case of an hyperelliptic curve, the involution is a symmetry in the construction of $W_1^{g}$, and one finds that the residues at $z_{\alpha}$ and $\overline{z}_{\alpha}$ are equal. Hence:
\bea
\partial_{t_j} \mathcal{F}(\vec{t}) & = & \sum_{g \geq 0} N^{2 - 2g} \partial_{t_j} F^{g}(\Sigma_{\infty}(\vec{t})) \nonumber \\
& = & - \sum_{\alpha} \Res_{z \rightarrow z_{\alpha}} \mathrm{d}x(z)\,2N\,W_1(x(z))\,\int^z_o \partial_{t_j} y\mathrm{d}x \nonumber \\
& = & \partial_{t_j}\ln\tau
\eea
This proves the identification of the resummation of symplectic invariants with the $\tau$ function.

\section{Refinement on the assumptions}
\label{sec:2222}
\subsection{A remark on hypothesis $(iii)$}
\label{sec:hypo3}

Starting from any solution to $\frac{1}{N}\partial_x \Psi = \mathbf{L} \Psi$, with $\mathbf{L}$ a given $2\times 2$ matrix, rational in $x$ and traceless, the connected correlators $\mathcal{W}_n$ were defined in \cite{BE09} and Section~\ref{sec:loopeqint} such that they satisfy "loop equations". When these loop equations are written in terms of the $\tau$ function, they appear identical to Virasoro constraints\footnote{Independently of \cite{BE09}, Virasoro constraints have also been found in \cite{KS09} in the case of $2 \times 2$ Schlesinger system.}. Loop equations can have many solutions, but their solution is unique if one requires some analyticity properties. In fact, provided there is a large parameter $N$ in the problem, and provided an expansion:
\beq
\label{eq:exp}\mathcal{W}_n = \sum_{g \geq 0} N^{2 - 2g - n}\,\mathcal{W}_n^{g}
\eeq
exists, the topological recursion gives the unique solution of the hierarchy of loop equations satisfied by $\mathcal{W}_n^g$, whose analytical properties are specified by the data of the spectral curve.

The matter is then, for the $\Psi$ one is interested in, to find the appropriate spectral curve $\Sigma_{\infty}$. It is often easy to prove that $\Psi$ has an expansion in $1/N$, and to discuss its analytical properties (see Section~\ref{sec:ana}). However, hypothesis $(iii)$ of Thm.~\ref{th2} is stronger. Indeed, the two essential properties of Eqn.~\ref{eq:exp} are:
\begin{itemize}
\item[$\bullet$] Fixed parity: only powers of $1/N$ with parity $(-1)^n$ appears in the expansion of $\mathcal{W}_n$.
\item[$\bullet$] Factorization property: the leading order of $\mathcal{W}_n$ is $O(N^{2 - n})$.
\end{itemize}
We believe that in practice, "fixing parity" is only a matter of rescaling correctly with $N$ the parameters of the differential system\footnote{This is not a void statement, because Theorem~\ref{th2} asks by convention for a $\frac{1}{N}$ in front of each $\partial_x$ in the differential system, and also that $\lim_{N \rightarrow \infty} \mathbf{L}$ exists.} $\partial_x \Psi = \mathbf{L}\Psi$ and that $\mathbf{L}^{(0)}$. In Appendix~\ref{sec:ctr}, we give an example based on a Lax pair associated to Painlev\'{e} II$_{\alpha}$ equation, where the fixed parity property does not hold if one does not rescale $\alpha$ as well.

Now, we argue that the factorization property is automatic, as soon as one realizes $\partial_x \Psi = \mathbf{L}\Psi$ as the member of a full integrable hierarchy. We present this argument for $2 \times 2$ systems with $1$ pole at $x = \infty$ (the case of interest in this article), but most of these ideas could be adapted without difficulty to several poles and larger systems. We are currently working out the generalization of all arguments presented in this article to the case of $d\times d$ differential systems \cite{BEenprepa}.

\subsection{Loop insertion operator and properties}

In some sector near $x = \infty$, we consider a solution $\Psi$ whose asymptotic is given by $\Psi = \widetilde{\Psi}_{\alpha}e^{N\,T \sigma_3}$, with $\widetilde{\Psi}$ regular at $x = \infty$, and:
\beq
\label{eq:asys}T = \sum_{j \geq 1} t_j\,x^j
\eeq
for some constants $t_{j}$. It is a standard result in integrable systems \cite{BBT} that $t_j$ can be used to define commuting flows. The result yields a matrix $\mathbf{L}$ and solution $\Psi$, which now depends on all $t_j$, with sectorwise asymptotic given by Eqn.~\ref{eq:asys}, and such that:
\bea
&& \frac{1}{N}\,\partial_x \Psi = \mathbf{L}\Psi \nonumber \\
&& \frac{1}{N}\,\partial_{t_j} \Psi = \mathbf{M}_j \Psi \nonumber \\
&& \mathbf{M}_j(x) = \left[\widetilde{\Psi}(x)\,x^j \mathbf{\sigma}_3\,\widetilde{\Psi}^{-1}(x)\right]_{-}
\eea
$\left[\cdots\right]_{-,x}$ denotes the divergent part when $x \rightarrow \infty$, and obviously $\mathbf{M}_j$ is a polynomial in $x$ of degree $j$. Let us define the \emph{loop insertion operator} as a Laurent series of operators when $x \rightarrow \infty$:
\beq
\label{eq:deg}\delta_{x} = \sum_{j \geq 1} \frac{\partial_{t_j}}{x^{j - 1}}
\eeq
It is merely a way to collect all the flows\footnote{$\frac{1}{N}\,\delta_x$ given by Eqn.~\ref{eq:deg} gives a realization of the formal loop insertion operator introduced in \cite{BE09}}.
\begin{lemma}
\label{thjjj}
We write $\widetilde{\cdots}$ the functions $\psi,\phi,\cdots$ with their essential singularity removed. The following $2 \times 2$ matrix:
\bea
\mathbf{P}(x) & \equiv & \frac{\mathbf{1} - \widetilde{\Psi}(x)\,\mathbf{\sigma}_3\,\widetilde{\Psi}^{-1}(x)}{2} \nonumber \\
\label{eq:PPP} & = & \left(\begin{array}{cc} -\overline{\psi}(x)\phi(x) & \widetilde{\psi}(x)\widetilde{\phi}(x) \\ - \widetilde{\overline{\psi}}(x)\widetilde{\overline{\phi}}(x) & \psi(x)\overline{\phi}(x)\end{array}\right)
\eea
is a projector, and we have:
\bea
\frac{1}{N}\,\delta_{x_2}\widetilde{\Psi}(x_1) & = & \frac{2(\mathbf{P}(x_2) - \mathbf{P}(x_1))}{x_2 - x_1}\,\widetilde{\Psi}(x_1) \\
\frac{1}{N}\,\delta_{x_2} \mathbf{P}(x_1) & = & \frac{[\mathbf{P}(x_1),\mathbf{P}(x_2)]}{x_2 - x_1}
\eea
\end{lemma}

\proof{\noindent Eqn.~\ref{eq:PPP} is a straightforward computation. We just notice, according to the representations of Eqns.~\ref{eq:says1}-\ref{eq:says2}, that the essential singularities cancels when we compute the cross-products: $\psi\overline{\phi} = \widetilde{\psi}\,\widetilde{\overline{\phi}}$ and $\overline{\psi}\phi = \widetilde{\overline{\psi}}\,\widetilde{\phi}$. Then, we observe that $\mathbf{P}$ is a rank $1$ matrix:
\beq
\mathbf{P} = {}^t\big(\widetilde{\phi}\quad\widetilde{\overline{\phi}}\big)\,\big(\widetilde{\psi}\quad\widetilde{\overline{\psi}}\big)\,\mathbf{J}
\eeq
with:
\beq
\mathbf{J} = \left(\begin{array}{cc} 0 & -1 \\ 1 & 0 \end{array}\right),\qquad \mathbf{J}^2 = -1,\qquad {}^t\mathbf{J} = \mathbf{J}^{-1} = - \mathbf{J}
\eeq
By definition $\mathrm{Tr}\,\mathbf{P} = 1$. Hence, $\mathbf{P}^2 = (\mathrm{Tr}\,\mathbf{P})\mathbf{P} = \mathbf{P}$.
Now, let us compute the action of the loop insertion operator on $\Psi$. $\partial_{t_j} \Psi = \mathbf{M}_j \Psi$ translates into:
\beq
\frac{1}{N}\,\partial_{t_j} \widetilde{\Psi}(x_1) = \left[\widetilde{\Psi}(x_1)\,x_1^j\mathbf{\sigma}_3\,\widetilde{\Psi}^{-1}(x_1)\right]_{-}\!\widetilde{\Psi}(x_1) - \widetilde{\Psi}(x_1)\,x^j\mathbf{\sigma}_3
\eeq
It is implied that $[\cdots]_-$ is the divergent part when $x_1 \rightarrow \infty$. Doing the summation over $j$:
\bea
\frac{1}{N}\,\delta_{x_2}\widetilde{\Psi}(x_1) & = & \left[\widetilde{\Psi}(x_1)\,\frac{\mathbf{\sigma}_3}{x_2 - x_1}\,\widetilde{\Psi}^{-1}(x_1)\right]_{-}\!\widetilde{\Psi}(x_1) - \widetilde{\Psi}(x_1)\,\frac{\mathbf{\sigma}_3}{x_2 - x_1} \nonumber \\
& = & \frac{\widetilde{\Psi}(x_1)\,\mathbf{\sigma}_3\,\widetilde{\Psi}^{-1}(x_1) - \widetilde{\Psi}(x_2)\,\mathbf{\sigma}_3\,\widetilde{\Psi}^{-1}(x_2)}{x_2 - x_1}\,\widetilde{\Psi}(x_1) \nonumber \\
& = & \frac{2(\mathbf{P}(x_2) - \mathbf{P}(x_1))}{x_2 - x_1}\,\widetilde{\Psi}(x_1) \nonumber
\eea
We apply this result to compute:
\bea
\frac{1}{N}\,\delta_{x_2}\mathbf{P}(x_1) & = & -\frac{1}{2} \delta_{x_2}\left( \widetilde{\Psi}(x_1)\,\mathbf{\sigma}_3\,\widetilde{\Psi}(x_1)^{-1}\right) \nonumber \\
& = & - \frac{\mathbf{P}(x_2) - \mathbf{P}(x_2)}{x_2 - x_1} \widetilde{\Psi}(x_1)\,\mathbf{\sigma}_3\,\widetilde{\Psi}^{-1}(x_1) + \widetilde{\Psi}(x_1)\,\mathbf{\sigma}_3\,\widetilde{\Psi}^{-1}(x_1) \frac{\mathbf{P}(x_2) - \mathbf{P}(x_1)}{x_2 - x_1} \nonumber \\
& = & \frac{-(\mathbf{P}(x_2) - \mathbf{P}(x_1))\mathbf{P}(x_1) + \mathbf{P}(x_1)(\mathbf{P}(x_2) - \mathbf{P}(x_1))}{x_2 - x_1} \nonumber \\
& = & \frac{[\mathbf{P}(x_1),\mathbf{P}(x_2)]}{x_2 - x_1} \nonumber
\eea}

\begin{corollary}
The integrable kernel is self-replicating:
\beq
\frac{1}{N}\,\delta_{x_2} \mathcal{K}(x_1,x_3) = - \mathcal{K}(x_1,x_2)\mathcal{K}(x_2,x_3)
\eeq
\end{corollary}

\begin{proposition}
\label{thyw}The correlators can be expressed in terms of the projector $\mathbf{P}$:
\bea
\mathcal{W}_1(x) & = & - N\,\mathrm{Tr}\,\mathbf{P}(x)\mathbf{L}(x) \\
\mathcal{W}_2(x_1,x_2) & = & -\frac{1}{2}\frac{\mathrm{Tr}\,\big(\mathbf{P}(x_1) - \mathbf{P}(x_2)\big)^2}{(x_1 - x_2)^2}
\eea
and for $n \geq 3$:
{\small \beq
\mathcal{W}_n(x_1,\ldots,x_n)  =  N^{2 - n}\,(-1)^{n + 1} \sum_{\sigma\,\mathrm{cycles}\,\mathrm{of}\,\mathfrak{S}_n} \frac{\mathrm{Tr}\,\mathbf{P}(x_1)\mathbf{P}(x_{\sigma(1)}) \cdots\mathbf{P}(x_{\sigma^{n - 1}(1)})}{(x_1 - x_{\sigma(1)})(x_{\sigma(1)} - x_{\sigma^2(1)})\cdots(x_{\sigma^{n - 1}(1)} - x_1)} \nonumber
\eeq}
Or concisely with the loop insertion operator:
\beq
\label{eq:lopin}\mathcal{W}_n(x_1,\ldots,x_n) = N^{-n}\,\delta_{x_1}\cdots\delta_{x_n} \ln \tau
\eeq
\end{proposition}
As a consequence of the existence of a loop insertion operator and Eqn.~\ref{eq:lopin}, the factorization property is automatic, and the requirements of hypothesis $(iii)$ have to be checked only on $\mathcal{W}_1(x)$.

\subsection{Analyticity properties of the $1/N$ expansion}
\label{sec:ana}

\subsubsection{Summary of the problem}

Let us start with two $2\times 2$ matrices $\mathbf{L}(x,t)$ and $\mathbf{M}(x,t)$ such that:
\begin{itemize}
\item[$\bullet$] $\mathbf{L}$ and $\mathbf{M}$ are traceless, and rational in $x$.
\item[$\bullet$] They are solution of the compatibility equation:
\beq
\frac{1}{N}\partial_t \mathbf{L} - \frac{1}{N}\partial_x \mathbf{M} + [\mathbf{L},\mathbf{M}] = 0
\eeq
\item[$\bullet$] They admit an asymptotic expansion when $N \rightarrow \infty$ of the form:
\beq
\mathbf{L} = \sum_{l \geq 0} N^{-l}\,\mathbf{L}^{(l)},\qquad \mathbf{M} = \sum_{l \geq 0} N^{-l}\,\mathbf{M}^{(l)}
\eeq
\end{itemize}
and consider the equations:
\beq
\label{eq:eqq}\frac{1}{N}\,\partial_x \Psi = \mathbf{L}\Psi, \qquad \frac{1}{N}\,\partial_t \Psi = \mathbf{M}\Psi
\eeq
In other words, we first restrict ourselves to a hierarchy with only one time $t$. We shall discuss the analytic properties of $\psi$ (the discussion would be similar for $\overline{\psi}$, $\phi$ and $\overline{\phi})$. From the differential system wrt $x$, we know (Section~\ref{sec:lN}) that it admits a large $N$ asymptotic expansion of the form:
\beq
\psi(x) = \left(\sum_{l \geq 0} N^{-l}\,\widetilde{\psi}^{(l)}(x)\right)e^{-N\int^x_o y(\xi)\mathrm{d}\xi}
\eeq
where $o$ is some integration constant which may depend on $t$. The leading order is given by:
\beq
\label{eq:sj}\widetilde{\psi}^{(0)} = \sqrt{\frac{\mathbf{L}_{21}}{2y}}
\eeq
and $y$ is determined by the large $N$ spectral curve associated to the differential system in $x$:
\beq
\Sigma_{\infty}\,:\qquad y^2 = - \mathrm{det}\,\mathbf{L}^{(0)}
\eeq
Similarly, we may define the large $N$ spectral curve associated to the differential system in $t$:
\beq
\underline{\Sigma}_{\infty}\,;\qquad \underline{y}^2 = - \mathrm{det}\,\mathbf{M}^{(0)}
\eeq
We shall see that $\Sigma_{\infty}$ and $\underline{\Sigma}_{\infty}$ play an important role to locate the singularities of $\psi$.

\subsubsection{Analytical properties}
\label{sec:qh}
Let us describe the analytical properties of $\widetilde{\psi}^{(0)}$ in the variable $x$:
\begin{itemize}
\item[$\bullet$] It has branchpoints (order $\frac{-(2r + 1)}{4}$, $r \in \mathbb{Z}$) at branchpoints of $\Sigma_{\infty}$.
\item[$\bullet$] It has branchpoints (order $\frac{(2r + 1)}{2}$, $r \in \mathbb{Z}$) at the zeroes or poles of $\frac{\mathbf{L}^{(0)}_{21}}{y}$ which are not branchpoints of $\Sigma_{\infty}$.
\end{itemize}
We can insert Eqn.~\ref{eq:sj} in one or the other differential system to determine recursively the subleading orders. It is more convenient to write the two second order equations satisfied by $\psi$:
\bea
\frac{1}{N^2}\,\partial_x^2 \psi + \frac{1}{N}\,B\partial_x \psi + V \psi & = & 0 \nonumber \\
\frac{1}{N^2}\,\partial_t^2 \psi + \frac{1}{N}\,\underline{B}\partial_t \psi + \underline{V} \psi & = & 0 \nonumber
\eea
where:
\bea
B & = & -\frac{\partial_x b}{b} = \sum_{l \geq 0} N^{-l}\,B^{(l)} \nonumber \\
V & = &\mathrm{det}\,\mathbf{L} - \frac{1}{N}\,\partial_x \mathbf{L}_{11} + \frac{1}{N}\,\mathbf{L}_{11}\frac{\partial_x \mathbf{L}_{21}}{\mathbf{L}_{21}} = -y^2 + \sum_{l \geq 1} N^{-l}\,V^{(l)} \nonumber \\
\underline{B} & = & -\frac{\partial_x \mathbf{M}_{21}}{\mathbf{M}_{21}} = \sum_{l \geq 0} N^{-l}\,\underline{B}^{(l)} \nonumber \\
\underline{V} & = & \mathrm{det}\,\mathbf{M} - \frac{1}{N}\,\partial_t \mathbf{M}_{11} + \frac{1}{N}\,\mathbf{M}_{11}\frac{\partial_t \mathbf{M}_{21}}{\mathbf{M}_{21}} = -\underline{y}^2 + \sum_{l \geq 1} N^{-l}\,\underline{V}^{(l)}
\eea
Knowing $\widetilde{\psi}^{(0)}$, we can compute recursively, if we assume that $B^{(0)}, \underline{B}^{(0)}$ are not identically zero:
\bea
\widetilde{\psi}^{(l + 1)} & = & \frac{1}{y\,B^{(0)}}\left[-\partial_x^2\widetilde{\psi}^{(l)} + (\partial_xy)\widetilde{\psi}^{(l)} + 2y(\partial_x\widetilde{\psi}^{(l)}) \right.\nonumber \\
\label{eq:kx} & & \left. \sum_{m = 0}^{l} B^{(l - m)}\,\partial_x \widetilde{\psi}^{(m)} - \big(V^{(l + 1 - m)} + y\,B^{(l + 1 - m)}\big)\widetilde{\psi}^{(m)}\right] \\
& = & \frac{1}{\underline{y}\,\underline{B}^{(0)}}\left[-\partial_t^2\widetilde{\psi}^{(l)} + (\partial_t\underline{y})\widetilde{\psi}^{(l)} + 2\underline{y}(\partial_t\widetilde{\psi}^{(l)}) \right.\nonumber \\
\label{eq:ky} & & \left. + \sum_{m = 0}^l \underline{B}^{(l - m)}\,\partial_t\widetilde{\psi}^{(m)} - \big(\underline{V}^{(l + 1 - m)} + \underline{y}\,\underline{B}^{(l + 1 - m)}\big)\right]
\eea
A priori, $\widetilde{\psi}^{(l + 1)}$ has singularities at points where $\widetilde{\psi}^{(m)}$ (for $0 \leq m \leq l$) had already singularities, and at poles of $\mathbf{L}$ and $\mathbf{M}$. Besides, if we consider the differential system wrt $x$, new singularities may appear at each step:
\begin{itemize}
\item[$\bullet$] At the branchpoints of $(\Sigma_{\infty})$ (through $1/y$).
\item[$\bullet$] At the zeroes (in the $x$ variable) of $B^{(0)}$.
\end{itemize}
If we rather consider the differential system wrt $t$, new singularities may appear only:
\begin{itemize}
\item[$\bullet$] At the branchpoints of $\underline{\Sigma}_{\infty}$ (through $1/\underline{y}$).
\item[$\bullet$] At the zeroes (in the $x$ variable) of $\underline{B}^{(0)}$.
\end{itemize}
Subsequently, $\psi^{(l)}(x)$ may have singularities only:
\begin{itemize}
\item[$\bullet$] At the poles in the expansion of $\mathbf{L}$ and $\mathbf{M}$.
\item[$\bullet$] At the branchpoints of $\Sigma_{\infty}$.
\item[$\bullet$] At the branchpoints of $\underline{\Sigma}_{\infty}$.
\item[$\bullet$] At the common zeroes (in the $x$ variable) of $B^{(0)}$ and $\underline{B}^{(0)}$.
\end{itemize}
In many examples (and in this article), the fourth point is the key ingredient to rule out singularities at the zeroes of $y$ which are not branchpoints. It is often enough to have one time $t$, but the same argument could be repeated in presence of other times $t_j$, giving a priori more restrictions on the position of the singularities. Eventually, the singularities of $\mathcal{W}_n$ can be inferred from their definition (Eqn.~\ref{eq:dee}) and the singularities of $\widetilde{\psi}^{(l)}$. We just notice that this definition does not introduce poles at coinciding points.

\subsection{Conclusion}
\label{sec:BJ}
We may formulate a stronger version of Theorem~\ref{th2}.
\begin{theorem}
\label{th3}
Assume that:
\begin{itemize}
\item[$(i)$] $\mathbf{L}$ depends on some parameter $N$, and has a limit when $N \rightarrow \infty$.
\item[$(ii)$] The spectral curve $\Sigma_N$ of the system Eqn.~\ref{eq:psipsi} has a large $N$ limit $\Sigma_{\infty}$ which is regular, and has genus $0$.
\item[$(iii)'$] $\mathcal{W}_1(x)$ admits an asymptotic expansion when $N \rightarrow \infty$ of the form : \mbox{$\mathcal{W}_1 = \sum_{g \geq 0} N^{1 - 2g}\,\mathcal{W}_1^{g}$}, and $\mathcal{W}_1^{g}(x)$ for $g \neq 0$ may have singularities only at branchpoints of $\Sigma_{\infty}$.
\end{itemize}
Then, $\mathcal{W}_n(x_1,\ldots,x_n)$ admits an expansion of the form:
\beq
\mathcal{W}_n = \sum_{g \geq 0} N^{2 - 2g - n}\,\mathcal{W}_n^{g}
\eeq
The expansion coefficients of the correlators have only singularities at the branchpoints of $\Sigma_{\infty}$ (for $2g-2+n>0$), and are computed by the topological recursion applied to $\Sigma_{\infty}$:
\beq
\mathcal{W}_n^{g}(x(z_1),\ldots,x(z_n))\mathrm{d}x(z_1)\cdots\mathrm{d}x(z_n) = \omega_n^{g}(\Sigma_{\infty})(z_1,\ldots,z_n)
\eeq
\end{theorem}
In practice, hypothesis $(iii)'$ requires to check that:
\begin{itemize}
\item[$\bullet$] $\mathcal{W}_1$ has an expansion in odd powers of $N$. We think that it can always be achieved in an appropriate rescaling of the parameters of $\mathbf{L}$ with $N$.
\item[$\bullet$] $\mathcal{W}_1^{g}$ has only singularities at branchpoints of $\Sigma_{\infty}$. This is not too difficult, because we know where singularities can arise in $\Psi$ (Section~\ref{sec:qh}).
\end{itemize}
A similar result holds when $\Sigma_{\infty}$ is not of genus $0$ under an extra hypothesis on $\mathcal{W}_1^{g}$, but is beyond of the scope of the present article.

\section{Application to the Tracy-Widom GUE law}

\subsection{Tracy-Widom GUE and integrability}
\label{sec:intp2}

The Tracy-Widom GUE law $\mathrm{F}_{\mathrm{GUE}}(s)$ can be defined as a Fredholm determinant:
\beq
\label{eq:AAA} \mathrm{F}_{\mathrm{GUE}}(s) = \mathrm{det}\left(\mathbf{1} - \mathbf{K}_{\mathrm{Ai}}\right)_{L^2(\left[s,\infty\right.\left[\right.)}
\eeq
where $\mathbf{K}_{\mathrm{Ai}}$ is the Airy kernel:
\beq
(\mathbf{K}_{\mathrm{Ai}}\cdot f)(x) = \int_{\mathbf{R}} \mathrm{d}y\,\frac{Ai(x)Ai'(y) - Ai'(x)Ai(y)}{x - y}\,f(y)
\eeq
In their celebrated article \cite{TW93}, Tracy and Widom have proved an alternative formula making the link with an integrable system:
\beq
\mathrm{F}_{\mathrm{GUE}}(s) = \exp\left(\int_{s}^{\infty}H(t)\,\mathrm{d}t\right)
\eeq
where $H(s) = -q^2(s)$, and $q(s)$ is the unique solution of the Painlev\'{e} II equation:
\beq
q''(s) = 2q^3(s) + sq(s)
\eeq
which satisfies \cite{HMC}:
\beq
q(s) \mathop{\sim}_{s \rightarrow +\infty} \mathrm{Ai}(s) \sim \frac{\exp(-\frac{2}{3}s^{3/2})}{2\sqrt{\pi}s^{1/4}}
\eeq
$H(s)$ can be identified with a Hamiltonian for PII \cite{Oka}, and $\mathrm{F}_{\mathrm{GUE}}(s)$ with a $\tau$-function associated to this family of Hamiltonians (\cite{FW00} in the sense of Okamoto, \cite{BD01} in the sense of Jimbo-Miwa-Ueno).

\vspace{0.2cm}

\noindent The Painlev\'{e} II equation appears \cite{FN80} as the compatibility condition of the following system for $\Psi(x,s)$:
\beq
\label{eq:Q} \left\{\begin{array}{rcl} \partial_x \Psi & = & \mathbf{L}\,\Psi \\ \partial_s \Psi & = & \mathbf{M}\,\Psi \\ \Psi & = & \widetilde{\Psi}\,\exp\left[i\left(\frac{4}{3}x^3 + sx\right)\mathbf{\sigma}_3\right] \quad \mathrm{when}\; x \rightarrow +\infty\end{array}\right.
\eeq
where $\widetilde{\Psi} = \mathbf{1} + O(1/x)$ when $x \rightarrow +\infty$. The Lax pair is given $(\mathbf{L},\mathbf{M})$ is given by:
\bea
\mathbf{L}(x,s) & = & \left(\begin{array}{cc} -4ix^2 -i(s + 2q^2(s)) & 4x q(s) + 2i p(s) \\ 4x q(s) - 2i p(s) & 4ix^2 + i(s + 2q^2(s))\end{array}\right) \nonumber \\
\mathbf{M}(x,s) & = & \left(\begin{array}{cc} - ix & q(s) \\ q(s) & ix \end{array}\right)
\eea
The necessary condition of existence of $\Psi$ is $\partial_{s} \mathbf{L} - \partial_{x} \mathbf{M} + [\mathbf{L},\mathbf{M}] = 0$. This implies that $q(s)$ is solution of PII, $p(s) = q'(s)$. The asymptotic behavior of $\Psi$ determines the asymptotic behavior of Eqn.~\ref{eq:Q} for $q(s)$, picking up the Hastings-McLeod solution of PII \cite{HMC}. The existence of the solution $\Psi(x,s)$ was shown in \cite{FN80}. This system is an example of isomonodromy problem, itself closely related to a Riemann-Hilbert problem. Many authors have contributed to the study of these systems, and we refer to \cite{FIKN} for a review of the theory.

\subsection{$s \rightarrow -\infty$ asymptotics and spectral curve}
\label{sec:AAA}
Let us introduce a redundant parameter $N$. We define:
\beq
\label{eq:ide}\left\{\begin{array}{c} x = N^{1/3}\,X \\ s = N^{2/3}\,S\end{array}\right.,\qquad q(s) = N^{1/3}Q(S)
\eeq
Then, Eqn.~\ref{eq:Q} is equivalent to:
\beq
\label{eq:Qm}  \left\{\begin{array}{rcl} \frac{1}{N}\,\partial_X\Psi & = & \mathbf{L}\,\Psi \\ \frac{1}{N}\,\partial_S \Psi & = & \mathbf{M}\,\Psi \\ \Psi & = & \widetilde{\Psi}\,\exp\left[iN\left(\frac{4}{3}X^3 + SX\right)\mathbf{\sigma}_3\right] \quad \mathrm{when}\; X \rightarrow +\infty\end{array}\right.
\eeq
with the Lax pair:
\bea
\mathbf{L}(X,S) & = & \left(\begin{array}{cc} -4iX^2 - i(S + 2Q^2(S)) & 4XQ(S) + \frac{2iQ'(S)}{N} \\ 4X Q(S) - \frac{2iQ'(S)}{N} & 4iX^2 +i(S + 2Q^2(S))\end{array}\right) \nonumber \\
\mathbf{M}(X,S) & = & \left(\begin{array}{cc} -iX & Q(S) \\ Q(S) & iX \end{array} \right)
\eea
and the compatibility equation:
\beq
\label{eq:eqQ} \frac{1}{N^2}Q''(S) = 2Q(S)^3 + SQ(S)
\eeq
We are now in the situation described in Section~\ref{sec:bas}, where each derivative appears with a prefactor $1/N$. Again, $Q(S)$ is given by the Hastings-McLeod solution to Painlev\'{e} II, which is the unique solution \cite{HMC} such that, for $S < 0$, $\lim_{N \rightarrow \infty} Q(S)^2 = - \frac{S}{2}$. Because of this property, we may study the $s \rightarrow -\infty$ asymptotics for the system Eqn.~\ref{eq:Q} by studying the $N \rightarrow \infty$ asymptotics of the system of Eqn.~\ref{eq:Qm}. Then, the basic remark is that the derivative terms are subleading.
After Eqn.~\ref{eq:sp}, the finite $N$ spectral curve for this system is:
\beq
Y^2 = -16X^4 - 8X^2S - S^2 - 4Q^4(S) - 4SQ^2(S) + \frac{4\big(Q'(S)\big)^2}{N^2}
\eeq
In the large $N$ limit, since $Q^2(S) \sim -S/2$ (we assume $S < 0$), we obtain:
\beq
\label{eq:sp2}(\Sigma_{\infty})\,;\qquad Y^2 = -16X^2\left(X^2 + \frac{S}{2}\right)
\eeq
It can be brought in a canonical form by rescaling:
\beq
(\widehat{\Sigma})\,:\qquad \widehat{y}^2 = \frac{1}{4}\widehat{x}^2(\widehat{x}^2 + 4)
\eeq
where
\beq
\widehat{x} = i \sqrt{\frac{8}{-S}}\,X,\qquad \widehat{y} = \frac{i}{S}\,Y
\eeq
For this transformation, $Y\mathrm{d}X = (-S/2)^{3/2}\,\widehat{y}\mathrm{d}\widehat{x}$. Going from the topological recursion of $\Sigma_{\infty}$ to that of $\widehat{\Sigma}$ is only a matter of rescaling:
\bea
F^{g}(\Sigma_{\infty}) & = & (-S/2)^{3(1 - g)}\,F^{g}(\widehat{\Sigma}) \\
\omega_n^{(g)}(\Sigma_{\infty}) & = & (-S/2)^{3(1 - g - n/2)}\,\omega_n^{g}(\widehat{\Sigma})
\eea
Thus:
\bea
\mathcal{F}(\Sigma_{\infty}) & = & \sum_{g \geq 0} N^{2 - 2g}\,F^{g}(\Sigma_{\infty}) \nonumber \\
& = & \sum_{g \geq 0} \big(N^{2/3}\,(-S/2)\big)^{3(1 - g)}\,F^{g}(\widehat{\Sigma}) \nonumber \\
\label{eq:fff}& = & \sum_{g \geq 0} (-s/2)^{3(1 - g)}\,F^{g}(\widehat{\Sigma})
\eea

\subsection{Existence of the $1/N$ expansion}
\label{sec:1N}
In our case, $\mathbf{L}(X,S)$ has only one pole, at $X = \infty$, which is already present in $\mathbf{L}^{(0)}(X,S)$. We define $Y(x)$ by Eqn.~\ref{eq:sp2} with the choice of the branch of the square root imposed by the $X \rightarrow \infty$ asymptotics in Eqn.~\ref{eq:Qm}:
\beq
Y(X) = -4iX\,\sqrt{X^2 + \frac{S}{2}},\qquad Y(x) \mathop{\sim}_{X \rightarrow +\infty} -4iX^2
\eeq
In order to identify $\mathcal{F}(\Sigma_{\infty})$ computed in Eqn.~\ref{eq:fff} with the $s \rightarrow -\infty$ asymptotic of the $\tau$ function of the initial system (Eqn~\ref{eq:Q}), we have to check that the assumptions announced in Sections~\ref{sec:loopeqint}-\ref{sec:N} hold.

\begin{remark}
$\Sigma_{\infty}$ is a regular genus $0$ spectral curve: it has two simple branchpoints, at $X = \pm \sqrt{-S/2}$. It admits a rational parametrization:
\beq
X(z) = \gamma\left(z + \frac{1}{z}\right),\qquad Y(z) = \frac{S}{2}\left(z^2 - \frac{1}{z^2}\right)
\eeq
where $\gamma = \sqrt{\frac{-S}{8}}$. Notice that it is not the spectral curve of a matrix model, since there is no $1/z$ term in $Y$. This is rather the curve of the limit of a matrix model.
\end{remark}

\begin{lemma}
\label{Lea}$\psi(X,S)$ (resp. $\overline{\psi}(X,S)$, $\phi(X,S)$, $\overline{\phi}(X,S)$) admits a $1/N$ expansion:
\beq
\psi(X,S) = \left(\widetilde{\psi}^{(0)}(X,S) +  \sum_{l \geq 1} N^{-l}\,\widetilde{\psi}^{(l)}(X,S)\right)\,\exp\left[iN\left(\frac{4}{3}X^3 + SX\right)\right]
\eeq
and for all $l \geq 1$, $\widetilde{\psi}^{(l)}(X,S)$ has only poles at $X = \pm \sqrt{\frac{-S}{2}}$ (the branchpoints of $\Sigma_{\infty}$). In particular, it does not have singularities at $X = 0$, the other zero of $Y$.
\end{lemma}

\proof{\noindent We already know that $Q^{(0)}(S) = \sqrt{\frac{-S}{2}}$, and Eqn.~\ref{eq:eqQ} implies that $Q$ has a $1/N^2$ expansion:
\beq
Q(S) = Q^{(0)}(S) + \sum_{l \geq 1} N^{-2l}\,Q^{(l)}(S) \nonumber
\eeq
Hence, $\mathbf{L}$ and $\mathbf{M}$ have a $1/N$ expansion. Let us apply the discussion of Section~\ref{sec:ana}, where $S$ plays the role of the time $t$. The large $N$ limit spectral curve $\Sigma_{\infty}$ associated to the system $\frac{1}{N}\,\partial_X \Psi = \mathbf{L}\Psi$ is:
\beq
Y(X) = \pm 4iX\sqrt{X^2 + S/2}
\eeq
and $\psi$ to leading order is given by:
\bea
\psi^{(0)}(X,S) & = & \mathrm{cte}^{(0)}(S)\,\sqrt{\frac{b^{(0)}(X,S)}{2Y(X)}}\,\exp\left[- N\int^X Y(\xi)\mathrm{d}\xi\right] \nonumber \\
& = & \mathrm{cte}^{(0)}(S)\,\left(\frac{-S/2}{X^2 + S/2}\right)^{1/4}\,\exp\left[\mp iN\left(\frac{4}{3}X^3 + SX\right)\right]
\eea
To agree with the asymptotic of Eqn.~\ref{eq:Qm}, we must choose the minus sign. On the other hand, the Iarge $N$ limit spectral curve $\underline{\Sigma}_{\infty}$ associated to the system $\frac{1}{N}\,\partial_S \Psi = \mathbf{M}\Psi$ is:
\beq
\underline{Y}(X) = \pm \sqrt{-\mathrm{det}\,\mathbf{M}^{(0)}} = \pm i \sqrt{X^2 + S/2}
\eeq
Hence, $Y(X)$ and $\underline{Y}(X)$ have no common zeroes (apart from the branchpoints $X = \pm \sqrt{-S/2}$). So, to all orders in $1/N$, $\widetilde{\psi}$ cannot have singularities at $X = 0$. The same discussion would hold for $\phi$, $\overline{\psi}$ and $\overline{\phi}$.}

\begin{corollary}
$\mathcal{W}_n(X_1,\ldots,X_n)$ admits a $1/N^2$ expansion:
\beq
\mathcal{W}_n(X_1,\ldots,X_n) = \sum_{g \geq 0} N^{2 - 2g - n}\,\mathcal{W}_n^{g}(X_1,\ldots,X_n)
\eeq
and the singularities of $\mathcal{W}_n^{g}(X_1,\ldots,X_n)$ away from $X_i = \infty$ are found only at branchpoints $X_i = \sqrt{-S/2}$.
\end{corollary}

\proof{\noindent There exists at least an expansion in $1/N$, and the position of the singularities at all orders is a consequence of Lemma~\ref{Lea}. We have seen in Thm.~\ref{thyw} that $\mathcal{W}_n$ starts as $O(N^{2 - n})$. It remains to prove that $\mathcal{W}_n$ has parity $(-1)^n$ in $N$. Let us stress the dependence in $N$ by writing $\mathbf{L}_N$, and $\Psi_N$ for the solution of Eqn.~\ref{eq:Qm}. We observe that ${}^t\mathbf{L}_{-N} = \mathbf{L}_N$, which implies that ${}^t\Psi^{-1}_{-N}$ obey the same differential system as $\Psi$. Moreover, ${}^t\Psi^{-1}_{-N}$ has the same asymptotic behavior near $x \rightarrow \infty$ as $\Psi_N$, and is also of determinant $1$. So, ${}^t\Psi_{-N}^{-1} = \Psi_N$, and at the level of the integrable kernel:
\bea
\mathcal{K}_N(x_1,x_2) & = & \frac{\psi_N(x_1)\overline{\phi}(x_2) - \overline{\psi}_N(x_1)\phi(x_2)}{x_1 - x_2} \nonumber \\
& = & \frac{\overline{\phi}_{-N}(x_1)\psi_{-N}(x_2) - \phi_{-N}(x_1)\overline{\psi}_{-N}(x_2)}{x_1 - x_2} \nonumber \\
& = & \mathcal{K}_{-N}(x_2,x_1) \nonumber
\eea
But, we can see on definition (Eqn.~\ref{eq:dee}) that the correlators take a $(-1)^n$ sign if we revert the orientation of all the cycles. Thus:
\beq
W_n(x_1,\ldots,x_n)_{-N} = (-1)^n W_n(x_1,\ldots,x_n)_{N}
\eeq}

Accordingly, we can apply Thm.~\ref{th2}, which says:
\bea
\mathcal{W}_n^{(g)}(X(z_1),\ldots,X(z_n))\mathrm{d}x(z_1)\cdots\mathrm{d}x(z_2) & = & \omega_n^{g}(\Sigma_{\infty})(z_1,\ldots,z_n) \nonumber \\
& = & (-S/2)^{3(1 - g - n/2)}\,\omega_n^{g}(\widehat{\Sigma})(z_1,\ldots,z_n) \nonumber \\
&&
\eea

\subsection{Tau-function and symplectic invariants}
\label{sec:tauN}
The next step is the computation of the $\tau$-function. Let us do the computation directly, as an illustration of the general proof given in Section~\ref{sec:N}. The $\tau$-function associated to the unique solution of the system satisfying Eqn.~\ref{eq:Qm} has:
\beq
\mathbf{T}_{\infty}(x,s) = iN\left(\frac{4}{3}X^3 + S X\right)\,\mathbf{\sigma}_3
\eeq
This is correct for any $N$, we see on this example that $\mathbf{T}_{\infty}$ is indeed given by its large $N$ limit. The $\tau$-function of Jimbo-Miwa-Ueno is given by (Eqn.~\ref{eq:tau0}):
\beq
\partial_{\sigma} \ln\tau =  2iN\,\Res_{X \rightarrow \infty} \mathrm{d}X\,X\mathcal{W}_1(X)
\eeq
Let us compare with the variation of $\mathcal{F}(\Sigma_{\infty})$ wrt $S$, given by Eqn.~\ref{eq:varF}. We have to represent $\partial_S Y\mathrm{d}X$ with the Bergman kernel:
\beq
B(z_1,z_2) = \frac{\mathrm{d}z_1\mathrm{d}z_2}{(z_1 - z_2)^2}
\eeq
We write:
\bea
(\partial_{S}Y\,\mathrm{d}X)(z) & = & \frac{-i\,X(z)\mathrm{d}X(z)}{\sqrt{X^2(z) + S/2}} = -i\,\mathrm{d}_z\left(\sqrt{X^2(z) + \frac{S}{2}}\right) \nonumber \\
& = & -i\,\Res_{\zeta \rightarrow z} B(z,\zeta)\,\sqrt{X^2(\zeta) + \frac{S}{2}} = -i\gamma\,\Res_{\zeta \rightarrow z} B(z,\zeta)\, \left(\zeta - \frac{1}{\zeta}\right) \nonumber \\
& = & i\gamma\,\Res_{\zeta \rightarrow 0,\infty} B(z,\zeta)\,\left(\zeta - \frac{1}{\zeta}\right) \nonumber
\eea
Thus:
\beq
\label{eq:RES}\partial_S F^g(\Sigma_{\infty}) = i\gamma\,\Res_{\zeta \rightarrow 0,\infty} \left(\zeta - \frac{1}{\zeta}\right) W_1^{g}(X(\zeta))\mathrm{d}X(\zeta)
\eeq
$\Sigma_{\infty}$ is an hyperelliptic curve, with involution $\zeta \mapsto \frac{1}{\zeta}$. From construction in the topological recursion, we know that:
\beq
\mathcal{W}_1^{g}(X(\zeta)) + \mathcal{W}_1^{g}(X(1/\zeta)) = (2y)_+(X(\zeta))
\eeq
where $y_+$ is the divergent part of $Y(X)$ when $X \rightarrow +\infty$:
\bea
y_+ & = & \left(-4iX\sqrt{X^2 + \frac{S}{2}}\right)_+ = -i(4X^2 + S) \nonumber \\
& = & \frac{iS}{2}\left(\zeta^2 + \frac{1}{\zeta^2}\right) \nonumber
\eea
With the last expression, one may check that adding $y_+$ to $W_1^g$ in Eqn.~\ref{eq:RES} do not change the result. In a first step, we may replace $W_1^{g}$ by:
\beq
\breve{\mathcal{W}}_1^{g} = \mathcal{W}_1^{g} - \delta_{g,0}y_+ \nonumber
\eeq
in Eqn.~\ref{eq:RES}. Now, we have the symmetry $\breve{\mathcal{W}}_1^g \rightarrow -\breve{\mathcal{W}}_1^{g}$ when $\zeta \rightarrow 1/\zeta$. This implies that the residues at $0$ and $\infty$ are equal:
\beq
\partial_S F^{g}(\Sigma_{\infty}) = 2i\gamma\,\Res_{\zeta \rightarrow \infty} \left(\zeta - \frac{1}{\zeta}\right) \breve{\mathcal{W}}_1^{g}(X(\zeta))\mathrm{d}X(\zeta) \nonumber
\eeq
In a second step, we can go back to $\mathcal{W}_1^g$:
\beq
\partial_S F^g(\Sigma_{\infty}) = 2i\gamma \Res_{\zeta \rightarrow \infty} \left(\zeta - \frac{1}{\zeta}\right) \mathcal{W}_1^{g}(\Sigma_{\infty})(X(\zeta))\mathrm{d}X(\zeta) \nonumber
\eeq
Another property of $\mathcal{W}_1^{g}$ from construction is that it behaves as $O(1/\zeta)$ (and even $O(1/\zeta^2)$ for $g \geq 1$) when $\zeta \rightarrow \infty$. Accordingly, we can replace $\left(\zeta - \frac{1}{\zeta}\right)$ by $\left(\zeta + \frac{1}{\zeta}\right)$. Eventually:
\bea
\partial_S F^g(\Sigma_{\infty}) & = & 2i\gamma\Res_{\zeta \rightarrow \infty}\left(\zeta + \frac{1}{\zeta}\right)W_1^g(\Sigma_{\infty})(X(\zeta))\mathrm{d}X(\zeta) \nonumber \\
& = & 2i\,\Res_{X \rightarrow \infty} \mathrm{d}X\,X\,\mathcal{W}_1^{g}(\Sigma_{\infty})X \nonumber
\eea
And, for the full series:
\beq
\partial_{S} \mathcal{F}(\Sigma_{\infty}) = 2iN\,\Res_{X \rightarrow \infty} \mathrm{d}X\,X\,\mathcal{W}_1(X) = \partial_S \ln \tau
\eeq

\subsection{Relation with the 1-matrix model with hard edge}
\label{sec:Con}
So far, we have proved that the Tracy-Widom GUE law $\mathrm{F}_{\mathrm{GUE}}(s)$, defined axiomatically from the Airy kernel (Eqn.~\ref{eq:AAA}), has the following $s \rightarrow -\infty$ expansion:
\beq
\mathrm{F}_{\mathrm{GUE}}(s) = C\,\exp\left(\sum_{g \geq 0} (-s/2)^{3(1 - g)}\,F^{g}(\widehat{\Sigma})\right)
\eeq
where $C$ is a constant, and $\widehat{\Sigma}$ is the spectral curve of equation $\widehat{y}^2 = \frac{1}{4}\widehat{x}^2(\widehat{x}^2 + 4)$. It admits the rational parametrization:
\beq
\left\{\begin{array}{l} x(z) = \frac{i}{2}(z + \frac{1}{z}) \\ y(z) = \frac{1}{2}\left(-z^2 + \frac{1}{z^2}\right) \end{array}\right.
\eeq
To complete the definition of the spectral curve, let us say that the physical sheet is associated to $x = +\infty$ ($z \rightarrow -i\infty$),  and that we take the Bergman kernel:
\beq
\widehat{B}(z_1,z_2) = \frac{\mathrm{d}z_1\mathrm{d}z_2}{(z_1 - z_2)^2}
\eeq

In \cite{BEMN}, starting from a matrix model with quadratic potential, where all eigenvalues are restricted to be smaller than a given $a$, we obtained heuristically the following $s \rightarrow -\infty$ expansion:
\beq
\label{eq:sj}\mathrm{F}_{\mathrm{GUE}}(s) = \mathrm{C}_{\mathrm{TW}}\,\exp\left(\sum_{g \geq 0} (-s/2)^{3(1 - g)}\,F^{g}(\Sigma_{\mathrm{TW}})\right)
\eeq
with a computable constant $\mathrm{C}_{\mathrm{TW}} = 2^{1/24}e^{\zeta'(-1)}$, and $\Sigma_{\mathrm{TW}}$ is the spectral curve of equation $y^2 = x + \frac{1}{x} - 2$. It admits the rational parametrization:
\beq
\left\{\begin{array}{l} x(\zeta) = \zeta^2 \\ y(\zeta) = \zeta - \frac{1}{\zeta} \end{array} \right.
\eeq
We associate the physical sheet to $\zeta \rightarrow +\infty$, and we take as Bergman kernel:
\beq
B_{\mathrm{TW}}(\zeta_1,\zeta_2) = \frac{\mathrm{d}\zeta_1\mathrm{d}\zeta_2}{(\zeta_1 - \zeta_2)^2}
\eeq

\begin{figure}[h!]
\begin{center}
\includegraphics[width=1.05\textwidth]{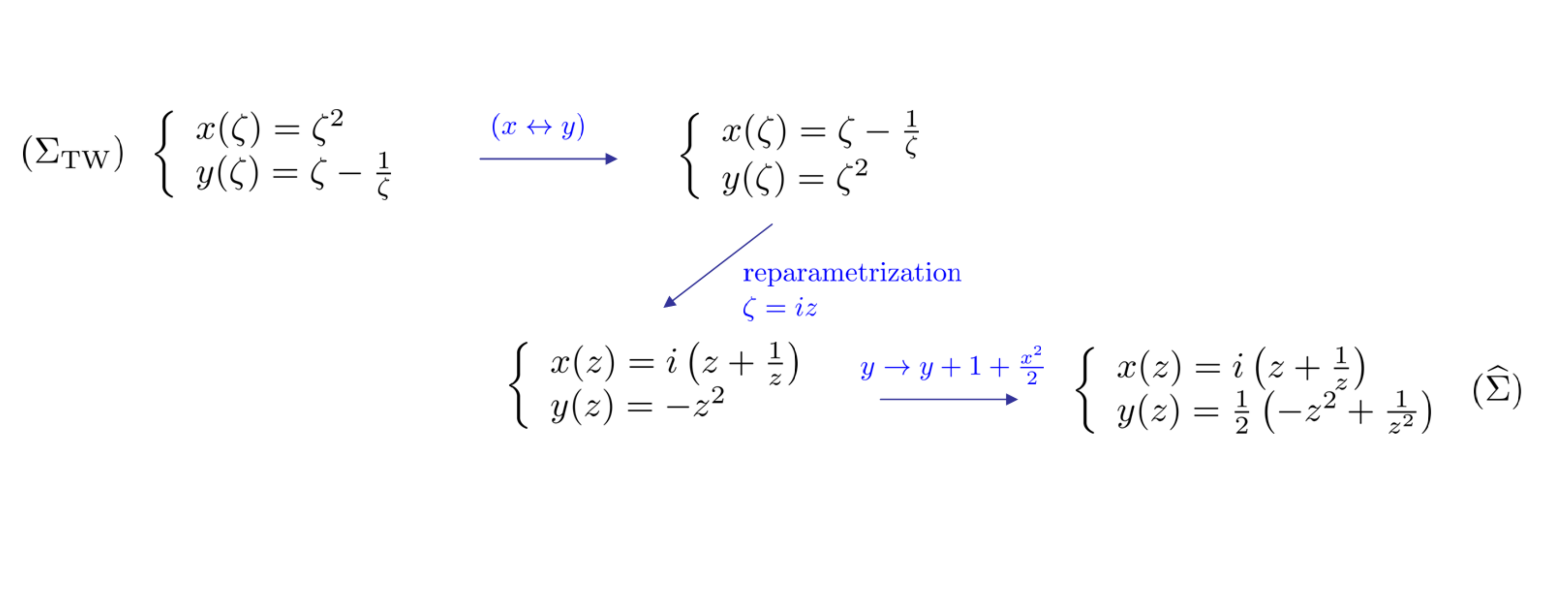}
\caption{\label{1} Symplectic equivalence of $\Sigma_{\mathrm{TW}}$ and $\widehat{\Sigma}$}
\end{center}
\end{figure}

Actually, these two spectral curves are related by a symplectic transformations, under which the $F^{g}$'s are invariant, as explained in Fig.~\ref{1}. Thus, we have justified Eqn.~\ref{eq:sj}, as announced in Prop~\ref{prop1}.
\vspace{0.4cm}

\subsection{Remark on correlation functions}

We insist on the fact that symplectic transformations do not conserve the correlation functions. A priori:
\beq
\omega_n^{g}(\Sigma_{\mathrm{TW}}) \neq \omega_n^{g}(\widehat{\Sigma})
\eeq
Nevertheless, one knows \cite{EORev} that their difference is an exact form in each variable.

It was argued in \cite{BEMN} that:
\beq
\omega_n(\Sigma_{\mathrm{TW}})(\zeta_1,\ldots,\zeta_n) \equiv \sum_{g \geq 0} (-s/2)^{3(1 - g - n/2)}\,\omega_n^{g}(\Sigma_{\mathrm{TW}})(\zeta_1,\ldots,\zeta_n)
\eeq
is the asymptotic expansion when $s \rightarrow -\infty$ of the (rescaled) correlation function of eigenvalues $\lambda_i = 2 + sN^{-2/3}(1 + \frac{\zeta_i^2}{2})$. In other words, for $s < 0$ and any Borel subsets $J_1,\ldots,J_n \subseteq \left.]-\infty,0[\right.$, we have:
\bea
& & \lim_{N \rightarrow \infty} \mathrm{Prob}\left[\lambda_1 \in J_1,\ldots,\lambda_n \in J_n\,|\,\lambda_{\mathrm{max}} \leq 2 + sN^{-2/3}\right] \nonumber \\
& = & \oint_{K_1}\frac{\mathrm{d}\zeta_1}{2i\pi}\cdots\oint_{K_n}\frac{\mathrm{d}\zeta_n}{2i\pi}\,\omega_n(\Sigma_{\mathrm{TW}})(\zeta_1,\ldots,\zeta_n)
\eea
in the sense of an asymptotic series when $s \rightarrow -\infty$. $K_i$ is the image on the curve of $J_i$.

A priori, this is different from the determinantal point process defined with the integrable kernel:
\beq
\mathcal{K}(x_1,x_2) = \frac{\psi(x_1)\overline{\phi}(x_2) - \overline{\psi}(x_1)\phi(x_2)}{x_1 - x_2}
\eeq
of the system of Eqn.~\ref{eq:Q}.

\newpage

\section{Conclusion}

\subsection*{Tracy-Widom law}

Let us summarize what we have obtained.

\begin{itemize}
\item[$\bullet$] Near an hard edge $a$ approaching the natural soft edge $a_{\rm max}$, the density of eigenvalues of a random hermitian matrix is (after appropriate rescaling) locally given  by
$$(\Sigma_{\mathrm{TW}})\,\, :\,\, y = \sqrt{x + \frac{1}{x} - 2}$$
and therefefore, the partition function in this regime is asymptotically given by the symplectic invariants of $\Sigma_{\mathrm{TW}}$:
$$
\ln Z(a) \sim \sum_g (1-a/a_{\rm max})^{(2-2g){3\over 2}}\,N^{2-2g}\,\, F^g(\Sigma_{\mathrm{TW}})
$$
and the eigenvalues correlation functions of the random hermitian matrix are also asymptotically given in this regime by the symplectic invariant correlators of $\Sigma_{\mathrm{TW}}$, i.e. by the $\omega_n^{(g)}(\Sigma_{\mathrm{TW}})$. Let us also mention that all this holds for any problem, solvable by symplectic invariants, whose spectral curve is locally given by $\Sigma_{\mathrm{TW}}$. This is for instance the case for the statistics of plane partitions near the boundary of the liquid region \cite{Eynpplane}.
\item[$\bullet$] The second step is that $\Sigma_{\mathrm{TW}}$ is symplectically equivalent to the spectral curve $\widehat{\Sigma}$:
$$(\widehat{\Sigma})\,\, :\,\, y = \frac{1}{4}x\sqrt{x^2 + 4},$$
which shows that
$$
\ln Z(a) \sim \sum_g (1-a/a_{\rm max})^{(2-2g){3\over 2}}\,N^{2-2g}\,\, F^g(\widehat{\Sigma}),
$$
and since $\widehat{\Sigma}$ is the classical spectral curve of a Lax pair for the Painlev\'e II equation, we recover, as expected, the large $s$ expansion of Tracy-Widom law.
\item[$\bullet$] However, in the symplectic invariance of $\Sigma_{\mathrm{TW}}$ and $\widehat{\Sigma}$, only the $F^g$'s are invariant, the correlators are not invariant.
\end{itemize}

It is not clear to us how the symplectic transformations (in particular $x \leftrightarrow y$, which is really the non trivial one at the level of the $F^g$) act on an integrable system of classical spectral curve $(x,y(x))$. It would be interesting to exhibit a full ($N$ dependent) integrable system whose classical spectral curve is directly $\Sigma_{\mathrm{TW}}$, without having to perform a symplectic transformation. Such a system should be related to PII, because the $\tau$ function is the same as that of Eqn.~\ref{eq:Q}, it can be expressed in terms of a solution of PII.

For $\beta$ ensembles, the generalization of this approach is an open problem. There exists a version of the topological recursion and symplectic invariants for all $\beta > 0$, computing the large $N$ expansion in the $\beta$ random matrix ensembles \cite{CE06} provided that the expansion exists. Yet, the analog of Section~\ref{sec:loopeqint} for $\beta \neq 2$ is missing: starting from an integrable system, it is not know how to define quantities $\mathcal{W}_n^{(\beta)}$ satisfying the loop equations of $\beta$ ensembles. One could expect that $\beta$ ensembles should be related to quantum integrable systems, having to do with Bethe ansatz, instead of classical Lax pair systems.

\subsection*{Loop equations and integrable systems}
We have reviewed that it is possible to associate loop equations (or Virasoro constraints) to any first order $2 \times 2$ rational differential integrable system. In fact, we have completed the work of \cite{BE09} by:
\begin{itemize}
\item[$\bullet$] Showing that hypothesis on analytical properties and $1/N$ expansion of the $n$-point functions need only to be checked for the $1$-point function.
\item[$\bullet$] Clarifying when these analytical properties and expansion properties holds.
\item[$\bullet$] Under all these assumptions, providing a proof of the former claim that the summation of symplectic invariants $F^{g}$ reconstruct the Tau function.
\end{itemize}
To fulfill this program, it was essential to include the differential system in an isomonodromic problem with a full set of times. It is in fact possible to associate loop equations to first order system of size $d \times d$, and we are currently working on the generalization of the same theory to these systems \cite{BEenprepa}.

\newpage

\setcounter{equation}{0}

\appendix

\section{A counterexample to the $1/N^2$ expansion}

\label{sec:ctr}
Consider a system $1/N\,\partial_x \Psi = \mathbf{L} \Psi$. The first loop equation is:
\beq
\mathcal{W}_2(x,x) + \big(\mathcal{W}_1(x)\big)^2 = N^2\,P_1(x)
\eeq
where $P_1(x)$ is a rational fraction given by:
\beq
P_1 = -\mathrm{det}\,\mathbf{L} = \frac{1}{2}\mathrm{Tr}\,\mathbf{L}^2
\eeq
Let us assume that $\mathbf{L} = \mathbf{L}^{(0)} + \frac{1}{N}\mathbf{L}^{(1)} + O(1/N^2)$. Then:
\beq
P_1 = \frac{1}{2} \mathrm{Tr}\,{\mathbf{L}^{(0)}}^2 + \frac{1}{N} \mathrm{Tr}\,\mathbf{L}^{(0)}\mathbf{L}^{(1)} + O(1/N^2)
\eeq
We want to find a case where the $\mathcal{W}_n$ do not have an expansion with fixed parity of $1/N$. It is enough to exhibit a system for which the first two terms in $P_1$ do not vanish.

We call Painlev\'{e} II$_{\alpha}$ the equation:
\beq
q''(s) = 2q(s)^3 + sq(s) - \alpha
\eeq
where $\alpha$ is a fixed parameter. This equation appears as the compatibility condition of the Lax system \cite{FN80}:
\bea
&& \left\{\begin{array}{l} \partial_x \Psi = \mathbf{L}\Psi \\ \partial_s \Psi = \mathbf{M}\Psi \end{array}\right. \nonumber \\
&& \mathbf{L}(x,s) = \left(\begin{array}{cc} -4ix^2 -i(s + 2q^2(s)) & 4xq(s) + 2iq'(s) + \frac{\alpha}{x} \\ 4xq(s) - 2iq'(s) + \frac{\alpha}{x} & 4ix^2 + i(s + 2q^2(s))\end{array}\right) \nonumber \\
&& \mathbf{M}(x,s) = \left(\begin{array}{cc} -ix & q(s) \\ q(s) & ix \end{array}\right) \nonumber
\eea
Let us do the rescaling:
\beq
\label{eq:ideA}\left\{\begin{array}{c} x = N^{1/3}\,X \\ s = N^{2/3}\,S\end{array}\right.,\qquad q(s) = N^{1/3}Q(S)
\eeq
This leads us to the compatibility equation:
\beq
\frac{1}{N^2}\,Q''(S) = 2Q(S)^3 + SQ(S) - \frac{\alpha}{N}
\eeq
and the Lax system:
\bea
&& \left\{\begin{array}{l} \frac{1}{N}\,\partial_X \Psi = \mathbf{L}\Psi \\ \frac{1}{N}\,\partial_S \Psi = \mathbf{M}\Psi\end{array}\right. \nonumber \\
&& \mathbf{L}(X,S) =  \left(\begin{array}{cc} -4iX^2 - i(S + Q^2(S)) & 4XQ(S) + \frac{2iQ'(S)}{N} + \frac{\alpha}{NX} \\ 4XQ(S) - \frac{2iQ'(S)}{N} + \frac{\alpha}{NX} & 4iX^2 + i(S + 2Q^2(S)) \end{array}\right) \nonumber \\
&& \mathbf{M}(X,S) = \left(\begin{array}{cc} -iX & Q(S) \\ Q(S) & iX \end{array}\right)\nonumber
\eea
When $N \rightarrow \infty$, we have to the first two orders:
\beq
Q(S) = \sqrt{-S/2} + \frac{1}{N}\,\frac{\alpha}{2S} + O(1/N^2) \nonumber
\eeq
We may compute:
\bea
& & \mathrm{Tr}(\mathbf{L}^{(0)}\mathbf{L}^{(1)}) \nonumber \\
& = & \mathrm{Tr} \left(\begin{array}{cc} -4iX^2 & 4X\sqrt{-S/2} \\ 4X\sqrt{-S/2} & 4iX^2\end{array}\right)\cdot\left(\begin{array}{cc} 0 & \frac{2X}{S} - \frac{i}{\sqrt{-2S}} + \frac{\alpha}{X} \\ \frac{2X}{S} + \frac{i}{\sqrt{-2S}} + \frac{\alpha}{X} & 0 \end{array}\right) \nonumber \\
& \neq & 0 \nonumber
\eea
Hence, the fixed parity property is not satisfied. Yet, one could also make the choice to rescale $\alpha$ in $A = \alpha/N$, and keep $A$ fixed. In this case, we could reproduce the argument of Section~\ref{sec:1N}, and check the fixed parity property. This counterexample supports the idea that a $1/N^2$ expansion may exist only in a good choice of rescaling with $N$ of the parameters of the differential system.

\section{Topological recursion and symplectic invariants}
\label{appB}

With the data of a spectral curve (see Section~\ref{sec:toporec}), the topological recursion defines a tower of differential forms as follows:
\bd
We define:
\beq
\omega_1^{(0)}(z) = -y(z)\,dx(z)
\eeq
\beq
\omega_2^{(0)}(z_1,z_2) = B(z_1,z_2)
\eeq
and by recursion on $2g-2+n$:
\bea
\omega_{n+1}^{(g)}(\Sigma)(z_0,\overbrace{z_1,\dots,z_n}^{J})
&=& \sum_i \Res_{z\to a_i}\, K(z_0,z)\,\Big[ \omega_{n+2}^{(g-1)}(\Sigma)(z,\bar z,J) \cr
&& +\sum_{h=0}^{g-1}\sum'_{I\subseteq J} \omega_{1+\# I}^{(h)}(\Sigma)(z,I)\,\omega_{1+n-\# I}^{(g-h)}(\Sigma)(\bar z,J\setminus I) \Big]
\eea
where the sum over $i$ is over all branchpoints $a_i$ (the zeroes of $dx(z)$), residues are taken as contour integrals on the Riemann surface ${\cal C}$, the kernel $K(z_0,z)$ is:
\beq
K(z_0,z) = -{\int_{z'=\bar z}^z \omega_2^{(0)}(\Sigma)(z_0,z')\over 2\,(\omega_1^{(0)}(\Sigma)(z)-\omega_1^{(0)}(\Sigma)(\bar z))}
\eeq
and in the sum $\sum_h\sum'_{I\subset J}$ the prime $'$ means that we exclude the pairs $(h=0,I=\emptyset)$ and $(h=g,I=J)$ from the sum.
\ed
An important property is that for $2g-2+n>0$, $\omega_n^{(g)}$ is a meromorphic form in each $z_i$, with poles only at branchpoints, and is symmetric in all $z_i$'s. For instance, we have:
\bea
\omega_1^{(1)}(z_0) & = & \sum_i \Res_{z\to a_i}\, K(z_0,z)\,B(z,\bar z) \\
\omega_3^{(0)}(z_0,z_1,z_2) & = & \sum_i \Res_{z\to a_i}\, K(z_0,z)\,(B(z,z_1)B(\bar z,z_2)+B(z,z_2)B(\bar z,z_1))
\eea

Then, having defined $\omega_n^{(g)}$ for $n\geq 1$:
\bd
For $g\geq 2$, we define the symplectic invariant $\omega_0^{(g)}(\Sigma)$ also denoted $F^g(\Sigma)$ as:
\beq
F^g(\Sigma)=\omega_0^{(g)}(\Sigma) = {1\over 2-2g}\,\sum_i \Res_{z\to a_i}\, \hat\Phi(z)\,\omega_1^{(g)}(\Sigma)(z)
\eeq
where
\beq
\hat \Phi(z) = \int_o^z\, y(z')\,dx(z')
\eeq
and the value of $F^g$ is independent of the base point $o$ and path used to compute the integral.
\ed
The definitions of $F^0(\Sigma)$ and $F^1(\Sigma)$ are more involved and we refer the reader to \cite{EORev}.


\newpage

\begin{thebibliography}{99}
\bibitem{BBT} O.~Babelon, D.~Bernard, M.~Talon, \emph{Introduction to classical integrable systems}, Cambridge Monographs on Mathematical Physics (2003)
\bibitem{BBDF07} J.~Baik, R.~Buckingham, J.~DiFranco, \emph{Asymptotics of Tracy-Widom distributions and the total integral of a Painlev\'{e} II function}, Comm. Math. Phys. \textbf{280} 2, pp 463-497 (2008), \href{http://arxiv.org/abs/0704.3636}{\texttt{math.FA/0704.3636}}
\bibitem{BE09} M.~Berg\`{e}re, B.~Eynard, \emph{Determinantal formulae and loop equations}, \href{http://arxiv.org/abs/0901.3273}{\texttt{math-ph/0901.3273}} (2009)
\bibitem{BE09l} M.~Berg\`{e}re, B.~Eynard, \emph{Universal scaling limits of matrix models, and $(p,q)$ Liouville gravity}, \href{http://arxiv.org/abs/0909.0854}{\texttt{math-ph/0909.0854}} (2009)
\bibitem{BI1999} P. Bleher, A. Its, \emph{Semiclassical asymptotics of orthogonal polynomials, Riemann-Hilbert
problem, and universality in the matrix model}, Ann. Math. \textbf{150} pp 185-266 (1999)
\bibitem{BD01} A.~Borodin, P.~Deift, \emph{Fredholm determinants, Jimbo-Miwa-Ueno tau-functions, and representation theory}, Comm. Pure App. Math.
\textbf{55} 9, pp 1160–1230 (2002), \href{http://arxiv.org/abs/math-ph/0111007}{\texttt{math-ph/0111007}}
\bibitem{BEMN} G.~Borot, B.~Eynard, S.N.~Majumdar, C.~Nadal, \emph{Large deviations of the maximal eigenvalue of random matrices}, submitted to J.Stat.Phys, \href{http://arxiv.org/abs/1009.1945}{\texttt{math-ph/1009.1945}} (2010)
\bibitem{BEenprepa} G.~Borot, B.~Eynard, \emph{Lax pairs and loop equations}, in preparation.
\bibitem{CM10} M.~Cafasso, O.~Marchal, \emph{Double scaling limits of random matrices and minimal $(2m,1)$ models: the merging of two cuts in a degenerate case}, \href{http://arxiv.org/abs/1002.3347}{\texttt{math-ph/1002.3347}} (2010)
\bibitem{CE06} L.~Chekhov, B.~Eynard, \emph{Matrix eigenvalue model: Feynman graph technique for all genera}, JHEP 0612:026 (2006), \href{http://arxiv.org/abs/math-ph/0604014}{\texttt{math-ph/0604014}}
\bibitem{DM06} D.S.~Dean, S.N.~Majumdar, \emph{Large deviations of extreme eigenvalues of random matrices}, Phys. Rev. Lett. \textbf{97}, pp 160-201 (2006), \href{http://arxiv.org/abs/cond-mat/0609651}{\texttt{cond-mat/0609651}}
\bibitem{DIK} P.~Deift, A.~Its, I.~Krasovsky, \emph{Asymptotics of the Airy-kernel determinant}, \href{http://arxiv.org/abs/math/0609451}{\texttt{math.FA/0609451}} (2006)
\bibitem{D1} P.~Deift, T.~Kriecherbauer, K.T-R.~McLaughlin, S.~Venakides, X.~Zhou, \emph{Asymptotics for polynomials orthogonal with
respect to varying exponential weights}, Int. Math. Res. Notices \textbf{16}, pp 759-782 (1997)
\bibitem{D2} P.~Deift, T.~Kriecherbauer, K.T-R.~McLaughlin, S.~Venakides, X.~Zhou, \emph{Strong asymptotics of orthogonal polynomials
with respect to exponential weights via Riemann–Hilbert techniques}, Comm. Pure Appl. Math. \textbf{52} 12, pp 1491-1552 (1999)
\bibitem{D3} P.~Deift, T.~Kriecherbauer, K.T-R.~McLaughlin, S.~Venakides, and X.~Zhou. \emph{Uniform asymptotics for polynomials orthogonal
with respect to varying exponential weights and applications to universality questions in random matrix theory}, Comm.
Pure Appl. Math. \textbf{52} 11, pp 1335-1425 (1999)
\bibitem{Erc} N.M.~Ercolani, K.T-R.~McLaughlin, \emph{Asymptotics of the partition function for random matrices via Riemann-Hilbert techniques, and applications to graphical enumeration}, Int. Math. Res. Not. \textbf{14}, pp 755-820 (2003), \href{http://arxiv.org/abs/math-ph/0211022}{\texttt{math-ph/0211022}}
\bibitem{Eyn2004} B.~Eynard, \emph{Topological expansion for the 1-hermitian matrix model correlation functions}, JHEP 024A:0904,  \href{http://arxiv.org/abs/hep-th/0407261}{\texttt{hep-th/0407261}}
\bibitem{EOFg} B.~Eynard, N.~Orantin, \emph{Invariants of algebraic curves and topological recursion}, \href{http://arxiv.org/abs/math-ph/0702045}{\texttt{math-ph/0702045}} (2007)
\bibitem{EORev} B.~Eynard, N.~Orantin, \emph{Algebraic methods in random matrices and enumerative geometry}, \href{http://arxiv.org/abs/0811.3531}{\texttt{math-ph/0811.3531}} (2008)
\bibitem{Eyn2008} B.~Eynard, \emph{Large N expansion of convergent matrix integrals, holomorphic anomalies, and background independence}, JHEP03(2009)003, \href{http://arxiv.org/abs/0802.1788}{\texttt{math-ph/0802.1788}}
\bibitem{Eynpplane} B.~Eynard, \emph{A matrix model for plane partitions}, J. Stat. Mech. 0910:P10011 (2009), \href{http://arxiv.org/abs/0905.0535}{\texttt{math-ph/0905.0535}}
\bibitem{FN80} H.~Flaschka, A.C.~Newell, \emph{Monodromy- and spectrum-preserving deformations, I}, Comm. Math. Phys. Volume \textbf{76} 1, pp 65-116 (1980), available \href{http://projecteuclid.org/DPubS?service=UI&version=1.0&verb=Display&handle=euclid.cmp/1103908189}{\texttt{here}}
\bibitem{FIKN} A.~Fokas, A.~Its, A.~Kapaev, V.~Novokshenov, \emph{Painlev\'{e} transcendents: the Riemann-Hilbert approach}, Mathematical Surveys and Monographs, Vol. 128 (2006)
\bibitem{FW00} P.J.~Forrester, N.S.~Witte, \emph{Application of the $\tau$-Function Theory of Painlevé Equations to Random Matrices: PIV, PII and the GUE}, Comm. Math. Phys. \textbf{219}, pp 357-398 (2001), \href{http://arxiv.org/abs/math-ph/0103025}{\texttt{math-ph/0103025}}
\bibitem{HMC} S.P.~Hastings, J.B.~McLeod, \emph{A boundary value problem associated with the second Painlev\'{e} transcendent and the Korteweg-deVries equation}, Archive for Rational Mechanics and Analysis, Volume \textbf{73}, Issue 1, pp 31-51, available \href{http://adsabs.harvard.edu/abs/1980ArRMA..73...31H}{\texttt{here}}
\bibitem{JM81} M.~Jimbo, T.~Miwa, K.~Ueno, \emph{Monodromy preserving deformation of linear ordinary differential equations with rational coefficients : I. General theory and $\tau$-function}, Physica D, Volume \textbf{2}, Issue 2, pp 306-352 (1981)
\bibitem{KS09} D.~Korotkin, H.~Samtleben, \emph{Generalization of Okamoto's equation to arbitrary $2 \times 2$ Schlesinger system}, Adv. Math. Phys. 461860 (2009) \href{http://arxiv.org/abs/0906.1962}{\texttt{nlin.SI/0906.1962}}
\bibitem{Mehtabook} M.L.~Mehta, {\em Random matrices}, 3rd edition, Pure and Applied Mathematics, Vol. 142, 3rd edn (Amsterdam: Elsevier/Academic)
\bibitem{Oka} K.~Okamoto, \emph{On the $\tau$-function of the Painlev\'{e} equations}, Physica D, Nonlinear Phenomena, Volume \textbf{2}, Issue 3,  pp 525-535 (1981)
\bibitem{RRV} J.A.~Ram\'{\i}rez, B.~Rider, B.~Vir\'{a}g, \emph{Beta ensembles, stochastic Airy spectrum, and a diffusion}, \href{http://arxiv.org/abs/math/0607331}{\texttt{math-ph/0607331}} (2006)
\bibitem{Scher} S.~Albeverio, L.~Pastur, M.~Shcherbina, \emph{On the $1/N$ expansion for some unitary invariant
ensembles of random matrices}, Comm. Math. Phys. \textbf{224}, pp 271-305 (2001), available \href{http://www.springerlink.com/content/llvrg62ykl477n6b/}{\texttt{here}}
\bibitem{TW93} C.~Tracy, H.~Widom, \emph{Level spacing distributions and the Airy kernel}, Comm. Math. Phys. \textbf{159} pp 151-174 (1994), \href{http://arxiv.org/abs/hep-th/9211141}{\texttt{hep-th/9211141}}


\end{thebibliography}
\end{document}